\documentclass[aip,jcp,preprint,longbibliography]{revtex4-1}
\usepackage{graphicx}
\usepackage{amsmath, amssymb}
\usepackage{color}
\usepackage{ulem}

\def\glv{\gamma_\mathrm{LV}}
\def\gsl{\gamma_\mathrm{SL}}
\def\gsv{\gamma_\mathrm{SV}}
\def\gs0{\gamma_\mathrm{S0}}
\def\gl0{\gamma_\mathrm{L0}}
\def\gv0{\gamma_\mathrm{V0}}
\def\Wsl{W_\mathrm{SL}}
\def\Wsv{W_\mathrm{SV}}
\def\ptl{\partial}
\def\ie{\textit{i.e.}, }
\def\Lztotal{L_{z}^\mathrm{total}}
\def\zpin{\zeta_\mathrm{pin}}
\def\dxizdz{\mathrm{d} \xi_{z}/\mathrm{d}z}
\def\xizcl{\xi_{z}^\mathrm{cl}}
\def\xiztop{\xi_{z}^\mathrm{top}}
\def\xizbot{\xi_{z}^\mathrm{bot}}
\def\xiztot{\xi_{z}^\mathrm{total}}
\def\Fzcl{F_{z}^\mathrm{cl}}
\def\Fztop{F_{z}^\mathrm{top}}
\def\Fzbot{F_{z}^\mathrm{bot}}
\def\eqref#1{(\ref{#1})}
\def\angb#1{\left<#1\right>}
\def\bm#1{\mbox{\boldmath $#1$}}
\def\xf{x_\mathrm{f}}
\def\yf{y_\mathrm{f}}
\def\zf{z_\mathrm{f}}
\def\xpf{x{'}_\mathrm{f}}
\def\ypf{y{'}_\mathrm{f}}
\def\zpf{z{'}_\mathrm{f}}
\def\dxf{\mathrm{d}x_\mathrm{f}}

\def\dzf{\mathrm{d}z_\mathrm{f}}

\def\dypf{\mathrm{d}y{'}_\mathrm{f}}

\def\zc{z_\mathrm{c}}
\def\xs{x_\mathrm{s}}
\def\ys{y_\mathrm{s}}
\def\zs{z_\mathrm{s}}
\def\xds{x{'}_\mathrm{s}}
\def\yds{y{'}_\mathrm{s}}
\def\zds{z{'}_\mathrm{s}}
\def\dzs{\mathrm{d}z_\mathrm{s}}

\def\zsv{z_\mathrm{SV}}
\def\zsl{z_\mathrm{SL}}
\def\zlblk{z_\mathrm{L}^\mathrm{blk}}
\def\zvblk{z_\mathrm{V}^\mathrm{blk}}
\def\xsf{x_\mathrm{SF}}
\def\xend{x_\mathrm{end}}
\def\zflr{z_\mathrm{flr}}
\def\zceil{z_\mathrm{ceil}}
\def\zpis{z_\mathrm{pis}}
\def\rnf{\rho_{V}^\mathrm{f}}
\def\rns{\rho_{A}^\mathrm{s}}
\def\usl{u_\mathrm{SL}}
\def\usv{u_\mathrm{SV}}
\def\usf{u_\mathrm{sf}}
\def\plblk{p_\mathrm{L}^\mathrm{blk}}
\def\pvblk{p_\mathrm{V}^\mathrm{blk}}
\def\rc{r_\mathrm{c}}
\def\drom{\mathrm{d}}
\newcommand{\osaka}{Department of Mechanical Engineering, Osaka University, 2-1 Yamadaoka, Suita 565-0871, Japan}
\newcommand{\tuswater}{Water Frontier Science \& Technology Research Center (W-FST),
Research Institute for Science \& Technology,
Tokyo University of Science,
1-3 Kagurazaka, Shinjuku-ku, Tokyo, 162-8601, Japan}
\begin{document}
\begin{flushright}
	The following article has been submitted to \textit{The Journal of Chemical Physics}.
\end{flushright}

\title{
Wilhelmy equation  revisited: a lightweight method to measure liquid-vapor, solid-liquid and solid-vapor interfacial tensions from a single molecular dynamics simulation
}

%
\author{Yuta Imaizumi}
\affiliation{\osaka}
\author{Takeshi Omori}%
\email{t.omori@mech.eng.osaka-u.ac.jp}
\affiliation{\osaka}
\author{Hiroki Kusudo}
\email{hiroki@nnfm.mech.eng.osaka-u.ac.jp}
\affiliation{\osaka}
\author{Carlos Bistafa}%
\email{bistafa@nnfm.mech.eng.osaka-u.ac.jp}
\affiliation{\osaka}
\author{Yasutaka Yamaguchi}
\email{yamaguchi@mech.eng.osaka-u.ac.jp}
\affiliation{\osaka}
\affiliation{\tuswater}
%
%
%
\date{\today}

\begin{abstract}
We have given theoretical expressions for the forces exerted on a so-called Wilhelmy plate, which we modeled as a quasi-2D flat and smooth solid plate immersed into a liquid pool of a simple liquid. All forces given by the theory, the local forces on the top, the contact line and the bottom of the plate as well as the total force, showed an excellent agreement with the MD simulation results. The force expressions were derived by a purely mechanical approach, which is exact and ensures the force balance on the control volumes arbitrarily set in the system, and are valid as long as the solid-liquid (SL) and solid-vapor (SV) interactions can be described by mean-fields.
In addition, we revealed that the local forces around 
the bottom and top of the solid plate can be related to the SL and SV interfacial tensions $\gsl$ and $\gsv$, 
and this was verified through the comparison with the 
SL and SV works of adhesion obtained by the thermodynamic integration (TI).
From these results, it has been confirmed that $\gsl$ and $\gsv$ as well as the liquid-vapor interfacial tension $\glv$ can be extracted from a single equilibrium MD simulation 
without the computationally-demanding  
calculation of the local stress distributions and the TI.
%
%
%
\end{abstract}


\maketitle 

%
\section{Introduction}
\label{sec:intro}
The behavior of the contact line (CL), where a liquid-vapor interface meets a solid surface, 
has long been a topic of interest in various scientific and engineering fields because it 
governs the wetting properties.~\cite{deGenne1985, Ono1960, Rowlinson1982, Schimmele2007, Drelich2019}
By introducing the concept of interfacial tensions and contact angle $\theta$,
Young's equation~\cite{Young1805} is given by
 \begin{equation}
   \gsl-\gsv+\glv \cos\theta = 0,
   \label{eq:Young}
 \end{equation}
where $\gsl$, $\gsv$ and $\glv$ denote
solid-liquid (SL), solid-vapor (SV) and liquid-vapor (LV) interfacial 
tensions, respectively. The contact angle is a common 
measure of wettability at the macroscopic scale.
Young's equation~\eqref{eq:Young} was first proposed based on the wall-tangential
force balance of interfacial tensions exerted on the CL in 1805 before the establishment 
of thermodynamics,~\cite{Gao2009} while recently it is often re-defined from a thermodynamic 
point of view instead of the mechanical force balance.~\cite{deGenne1985} 
\par
Wetting is critical especially in the nanoscale with a large 
surface to volume ratio, \textit{e.g.,} in the fabrication 
process of semiconductors,~\cite{Tanaka1993} 
where the length scale of the structure has reached 
down to several nanometers. 
%
%
From a microscopic point of view, \citet{Kirkwood1949} first provided the theoretical framework of surface tension based on the statistical mechanics, and molecular dynamics (MD)  
and Monte Carlo (MC) 
simulations have been carried out for the microscopic understanding of wetting through the connection  
with the interfacial tensions.~\cite{Nijmeijer1990_theor, Nijmeijer1990_simul, Tang1995, Gloor2005, Ingebrigtsen2007, Das2010, Weijs2011, Seveno2013, Surblys2014, Nishida2014, Lau2015, Yamaguchi2019, Kusudo2019, Bey2020, Grzelak2008, Leroy2009, Leroy2010, Kumar2014, Leroy2015, Ardham2015, Kanduc2017, Kanduc2017a, Jiang2017, Surblys2018, Ravipati2018}
Most of these works on a simple flat and smooth solid surface 
indicated that 
the apparent contact angle of the meniscus or droplet obtained in the 
simulations corresponded well to the one predicted by Young's equation~\eqref{eq:Young} 
using the interfacial tensions calculated through a mechanical manner and/or
a thermodynamic manner, where Bakker's equation and extended one 
about the relation between stress distribution around LV, SL or SV interface and 
corresponding interfacial tension have played a key role.~\cite{Yamaguchi2019}
On the other hand, on inhomogeneous or rough surfaces, 
the apparent contact angle did not seem to correspond well to the predicted 
one,~\cite{Leroy2010, Giacomello2016,Zhang2015,Zhang2019} because the pinning force 
exerted from the solid must be included in the wall-tangential force balance.~\cite{Kusudo2019}
\par
The Wilhelmy method~\cite{Wilhelmy1863} has been applied as one of 
the most common methods to experimentally measure the LV interfacial tension, \ie surface tension,  
or the contact angle.~\cite{Volpe2018}
In this method, the force on a solid sample vertically immersed 
into a liquid pool is expressed from the force balance by
\begin{equation}
  \Lztotal = 
  l \glv \cos \theta + mg - \rho gV,
\label{eq:Wilhelmy_full}
\end{equation}
where $\Lztotal$ is the total downward force (load) measured on the sample, the contact angle $\theta$ is defined on the liquid side, $l$ is the CL perimeter, $m$ is the sample mass, $V$ 
denotes the volume of the sample immersed in a liquid
of density $\rho$, and $g$ stands for the acceleration 
of gravity. 
The history of the Wilhelmy method and practical issues 
mainly from a macroscopic point of view are well summarized 
in a review article.~\cite{Volpe2018}
%
In the nanoscale, the gravitational force and buoyancy
respectively as the 2nd and 3rd terms on the RHS of
Eq.~\eqref{eq:Wilhelmy_full} are negligible, and it follows that 
\begin{equation}
  \xiztot \approx \glv \cos\theta,
  \label{eq:Wilhelmy}
\end{equation}
where the force per CL length $\xiztot$ is defined by
\begin{equation}
  \xiztot \equiv \frac{\Lztotal}{l}.
  \label{eq:def_xiztot}
\end{equation}
From Eq.~\eqref{eq:Wilhelmy},
one can estimate unknown $\glv$ from $\xiztot$ 
and $\theta$ determined by the apparent meniscus shape, 
or unknown $\theta$ from $\xiztot$ and 
$\glv$ as a known physical property.
Apparently, the sign of $\xiztot$
is directly related to the wettability, \ie the force is downward 
for a wettable solid sample with $\theta < \pi/2$.
\par
It is often modeled, typically with a macroscopic 
schematic illustrating the balance of forces acting on the 
solid sample, as if the solid sample is `pulled' locally 
at the CL toward the direction tangential to the LV 
interface. In such a model, the wall-tangential component of 
this force $l \glv \cos \theta$ in Eq.~\eqref{eq:Wilhelmy_full}
seems to act on the solid locally at the CL; 
however, 
it is not correct from a microscopic 
point of view\cite{Marchand2012,Das2011,Weijs2013}. As a straightforward example, consider the case with $\theta = \pi/2$: such model claims that the local wall-tangential force from the fluid around the CL must be zero because $\cos \theta=0$, whereas the fluid density $\rho$ along the wall-tangential direction $z$ changes with $\ptl \rho/\ptl z \ne 0$ around the CL, which should form an inhomogeneous force field for the solid in the $z$-direction.
Probably due to the difficulty of the direct 
experimental measurement, few studies have been carried out 
specifically about the local force on the solid in comparison with 
Young's equation so far. 
Among them, 
Das et al.\cite{Das2011} and Weijs et al.\cite{Weijs2013} proposed a model that describes the local force on the solid around the CL per unit length as $\glv(1+\cos \theta)$, which was based on the density functional theory with the sharp kink approximation.~\cite{Merchant1992,Getta1998}
This model was later examined by MD simulations for a simple liquid.~\cite{Seveno2013}
%
\par
In this work, 
we revisited the forces exerted on the Wilhelmy plate with non-zero thickness and derived theoretical expressions of the local forces on the CL and on the top and bottom of the plate as well as the total force on the plate. The derivations were done by a purely mechanical approach, which ensured the force balance on the arbitrarily set control volumes, and the connection to the thermodynamics was given by the extended Bakker equation.~\cite{Yamaguchi2019} We also verified the present theoretical results by MD simulations. 
%
As a major outcome of the expressions of the local forces, we will show in this article that all the interfacial tensions involved in the system, $\glv$, $\gsl$ and $\gsv$, can be measured from a single equilibrium MD simulation without computationally-demanding calculations. 
\section{Method}
\label{sec:method}
\subsection{MD Simulation}
\begin{figure}
  \begin{center}
    \includegraphics[width=0.9\linewidth]{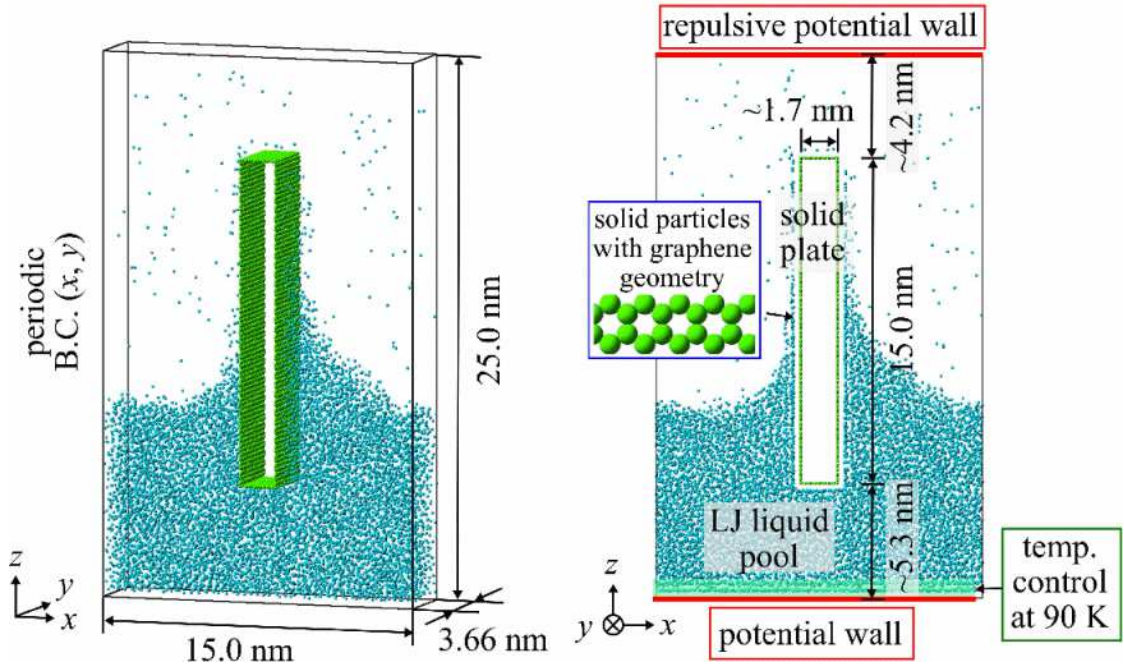}
  \end{center} 
  \caption{\label{Fig:system}
Equilibrium molecular dynamics (MD) simulation systems of a quasi-2D meniscus 
formed on a hollow rectangular solid plate dipped into a liquid pool of a 
simple Lennard-Jones (LJ) fluid: the Wilhelmy MD system.
}
\end{figure}
We employed equilibrium MD simulation systems of a quasi-2D 
meniscus formed on a hollow rectangular solid plate (denote by 
`solid plate' hereafter) dipped into a liquid pool of a simple 
fluid as shown in Fig.~\ref{Fig:system}. 
We call this system the `Wilhelmy MD system' hereafter.
Generic particles interacting through a LJ potential 
were adopted as the fluid particles. The 12-6 LJ potential 
given by
\begin{equation}
   \Phi^\mathrm{LJ}(r_{ij}) = 
  4\epsilon \left[
    \left(\frac{\sigma}{r_{ij}}\right)^{12} 
    -
    \left(\frac{\sigma}{r_{ij}}\right)^{6} 
    +
    c_{2}^\mathrm{LJ}\left(\frac{r_{ij}}{\rc}\right)^2 
    +
    c_{0}^\mathrm{LJ}
  \right],
  \label{eq:LJ}
\end{equation}
was used for the interaction between fluid particles, 
where $r_{ij}$ is the distance between the particles $i$ at position 
$\bm{r}_{i}$ and $j$ at $\bm{r}_{j}$, while $\epsilon$ and $\sigma$ denote the LJ energy and length parameters, respectively. This LJ interaction was truncated at a cut-off distance of $\rc=3.5 \sigma$ and quadratic functions were added so that the potential and interaction force smoothly vanished at $\rc$.
The constant values of
$c_{2}^\mathrm{LJ}$ and $c_{0}^\mathrm{LJ}$
were given in our previous study.~\cite{Nishida2014}
Hereafter, fluid and solid particles are denoted by `f' and `s', respectively and corresponding combinations are indicated by subscripts.
\par
A rectangular solid plate in contact with the fluid was prepared  
by bending a honeycomb graphene sheet, where the solid particles 
were fixed on the coordinate with the positions of 2D-hexagonal 
periodic structure with an inter-particle distance $r_\mathrm{ss}$ 
of 0.141~nm. The zigzag edge of the honeycomb structure 
was set parallel to the $y$-direction with locating solid 
particles at the edge to match the hexagonal periodicity.
The right and left faces were set at $x=\pm x_\mathrm{s}$ parallel
to the $yz$-plane, and the top and bottom faces were parallel 
to the $xy$-plane. Note that the distance between the left and 
right faces $2x_\mathrm{s}\approx 1.7$~nm was larger than the 
cutoff distance $\rc$.
\par
The solid-fluid (SF) interaction, which denotes SL or SV interaction, was also expressed by the LJ potential in Eq.~\eqref{eq:LJ}, where the length parameter $\sigma_\mathrm{sf}$ was given by the Lorentz mixing rule, while the energy parameter $\epsilon_\mathrm{sf}$ was changed in a parametric manner by multiplying a SF interaction coefficient $\eta$ to the base value $\epsilon^{0}_\mathrm{sf}=\sqrt{\epsilon_\mathrm{ff}\epsilon_\mathrm{ss}}$ as
\begin{equation}
\label{eq:def_eta}
\epsilon_\mathrm{sf} = \eta \epsilon^{0}_\mathrm{sf}.
\end{equation}
%
This parameter $\eta$ expressed the wettability, \ie $\eta$ and the contact angle of a hemi-cylindrically shaped equilibrium droplet on a homogeneous flat solid surface had a one-to-one correspondence~\cite{Nishida2014, Yamaguchi2019, Kusudo2019}, and 
we set the parameter $\eta$ between 0.03 and 0.15 so that 
the corresponding cosine of the contact angle $\cos \theta$ 
may be from $-0.9$ to $0.9$. The definition of the 
contact angle is described later in Sec.~\ref{sec:resdis}.
Note that due to the fact that the solid-solid inter-particle distance 
$r_\mathrm{ss}$ shown in  Table~\ref{tab:table1}
were relatively small compared to  the LJ length parameters 
$\sigma_\mathrm{ff}$ and $\sigma_\mathrm{fs}$, the surface is considered 
to be very smooth, and the wall-tangential force from the solid on the 
fluid, which induces pinning of the CL, is 
negligible.~\cite{Yamaguchi2019, Kusudo2019}
\par
In addition to these intermolecular potentials, we set a horizontal 
potential wall on the bottom (floor) of the calculation cell 
fixed at $z=\zflr$ about 5.3~nm below the bottom of the solid plate, which interacted only with 
the fluid particles with a one-dimensional potential field $\Phi_\mathrm{flr}^\mathrm{1D}$
as the function of the distance from the wall given by
\begin{equation}
  \label{eq:potentialbath}
  \Phi_\mathrm{flr}^\mathrm{1D}(z'_{i})=
  4\pi \rho_{n} \epsilon^{0}_\mathrm{sf}
  \sigma_\mathrm{sf}^{2}
  \left [ 
    \frac{1}{5} \left(
      \frac{\sigma_\mathrm{sf} }{z'_{i} }
    \right)^{10}
    \!\!\!\! - 
    \frac{1}{2} \left(
      \frac{\sigma_\mathrm{sf} }{z'_{i} }
    \right)^{4}
      +
      c_{2}^\mathrm{flr}
      \left(\frac{z'_{i}}{\zc^\mathrm{flr}}\right)^2 
      +
      c_{1}^\mathrm{flr}
      \left(\frac{z'_{i}}{\zc^\mathrm{flr}}\right)
    +
    c_{0}^\mathrm{flr}
 \right],
 \quad z'_{i}\equiv z_{i} - z_\mathrm{flr}
\end{equation}
where $z_{i}$ is the $z$-position of fluid particle $i$.
This potential wall mimicked a mean potential field created 
by a single layer of solid particles with a uniform area 
number density $\rho_{n}$. 
Similar to Eq.~\eqref{eq:LJ}, this potential field 
in Eq.~\eqref{eq:potentialbath} was 
truncated at a cut-off distance of 
$\zc^\mathrm{flr}=3.5 \sigma_\mathrm{sf}$ and a quadratic function was added so that the potential and interaction 
force smoothly vanished at $\zc^\mathrm{flr}$. 
As shown in Fig.~\ref{Fig:system}, fluid particles 
were rather strongly attracted on this plane because this 
roughly corresponded to a solid wall showing complete wetting.
With this setup, the liquid pool was stably kept even 
when the liquid pressure is low with a highly wettable solid plate.
Furthermore, we set another horizontal 
potential wall on the top (ceiling) of the calculation 
cell fixed at $z=\zceil$ about 4.7~nm above the top of the 
solid plate exerting a repulsive potential field $\Phi_\mathrm{ceil}^\mathrm{1D}$ 
on the fluid particles given by
\begin{equation}
  \label{eq:potentialbath_top}
  \Phi_\mathrm{ceil}^\mathrm{1D}(z''_{i})=
  4\pi \rho_{n} \epsilon^{0}_\mathrm{sf}
  \sigma_\mathrm{sf}^{2}
  \left [ 
    \frac{1}{5} \left(
      \frac{\sigma_\mathrm{sf} }{z''_{i} }
    \right)^{10}
      +
      c_{2}^\mathrm{ceil}
      \left(\frac{z''_{i}}{\zc^\mathrm{ceil}}\right)^2 
    +
      c_{1}^\mathrm{ceil}
      \left(\frac{z''_{i}}{\zc^\mathrm{ceil}}\right)
    +
    c_{0}^\mathrm{ceil}
 \right],
 \quad z''_{i}\equiv \zceil - z_{i},
\end{equation}
where a cut-off distance of 
$\zc^\mathrm{ceil}=
\sigma_\mathrm{sf}$ was set to
express a repulsive potential wall.
\par
The periodic boundary condition 
was set in the horizontal $x$- and $y$-directions, where the 
system size in the $y$-direction $l_{y}\approx3.66$~nm matched 
the hexagonal periodicity of the graphene sheet. 
The temperature of the system was maintained at a constant temperature 
of $T_\mathrm{c}$ at 90~K, which was above the triple point 
temperature,~\cite{Mastny2007} by velocity rescaling applied to the 
fluid particles within 0.8~nm from the floor wall regarding the 
velocity components in the $x$- and $y$-directions. Note that this 
region was sufficiently away from the bottom of the solid plate and 
no direct thermostating was imposed on around the solid plate, so 
that this temperature control had no effects on the present results. 
\par
With this setting, a quasi-2D LJ liquid of a meniscus-shaped LV interface 
with the CL parallel to the $y-$direction was formed as an equilibrium 
state as exemplified in Fig.~\ref{Fig:system}, where a liquid bulk with 
an isotropic density distribution existed above the bottom wall by 
choosing a proper number of fluid particles $N_\mathrm{f}$ as shown 
in Fig.~\ref{Fig:distribtution}. 
We checked that the temperature was constant in the whole system after 
the equilibration run described below.
Note also that in the present quasi-2D systems, effects of the CL curvature 
can be neglected.~\cite{Boruvka1977, Marmur1997line, Ingebrigtsen2007, Leroy2010, Weijs2011, Nishida2014, Yamaguchi2019,Kusudo2019} 
The velocity Verlet method was applied for the integration of the Newtonian equation of motion with a time increment of 5~fs for all systems. The simulation parameters are summarized in Table~\ref{tab:table1} with the corresponding non-dimensional ones, which are normalized by the 
corresponding standard values based on $\epsilon_\mathrm{ff}$,
$\sigma_\mathrm{ff}$ and $m_\mathrm{f}$. 
%
\par
The physical properties of each equilibrium system with various 
$\eta$ values were calculated as the time average of 40~ns, 
which followed an equilibration run of more than 10~ns.
\begin{table*}[!t]
\caption{\label{tab:table1} 
Simulation parameters and their corresponding non-dimensional values.
}
\begin{ruledtabular}
\begin{tabular}{cccc}
property  & value & unit & non-dim. value
\\ \hline
$\sigma_\mathrm{ff}$ & 0.340 & nm & 1
\\
$\sigma_\mathrm{sf}$ &  0.357 & nm & 1.05
\\
$\epsilon_\mathrm{ff}$ & $1.67 \times 10^{-21}$ & J & 1
\\
$\epsilon^{0}_\mathrm{sf}$
& $1.96\times 10^{-21}$ & J & 1.18
\\
$\epsilon_\mathrm{sf}$ & 
$\eta \times \epsilon^{0}_\mathrm{sf}$
\\
$\eta$ &
0.03 -- 0.15 & - & -
\\
$m_\mathrm{f}$ & $6.64 \times 10^{-26}$ & kg & 1
\\
$T_\mathrm{c}$ & 90  & K & 0.703
\\
$N_\mathrm{f}$ & 10000 - 15000  & - & -
\end{tabular}
\end{ruledtabular}
\end{table*}
\section{Results and discussion
\label{sec:resdis}}
\label{sec:result}
\subsection{Contact angle and force on the solid plate}
\begin{figure}
\centering
\includegraphics[width=0.9\linewidth]
{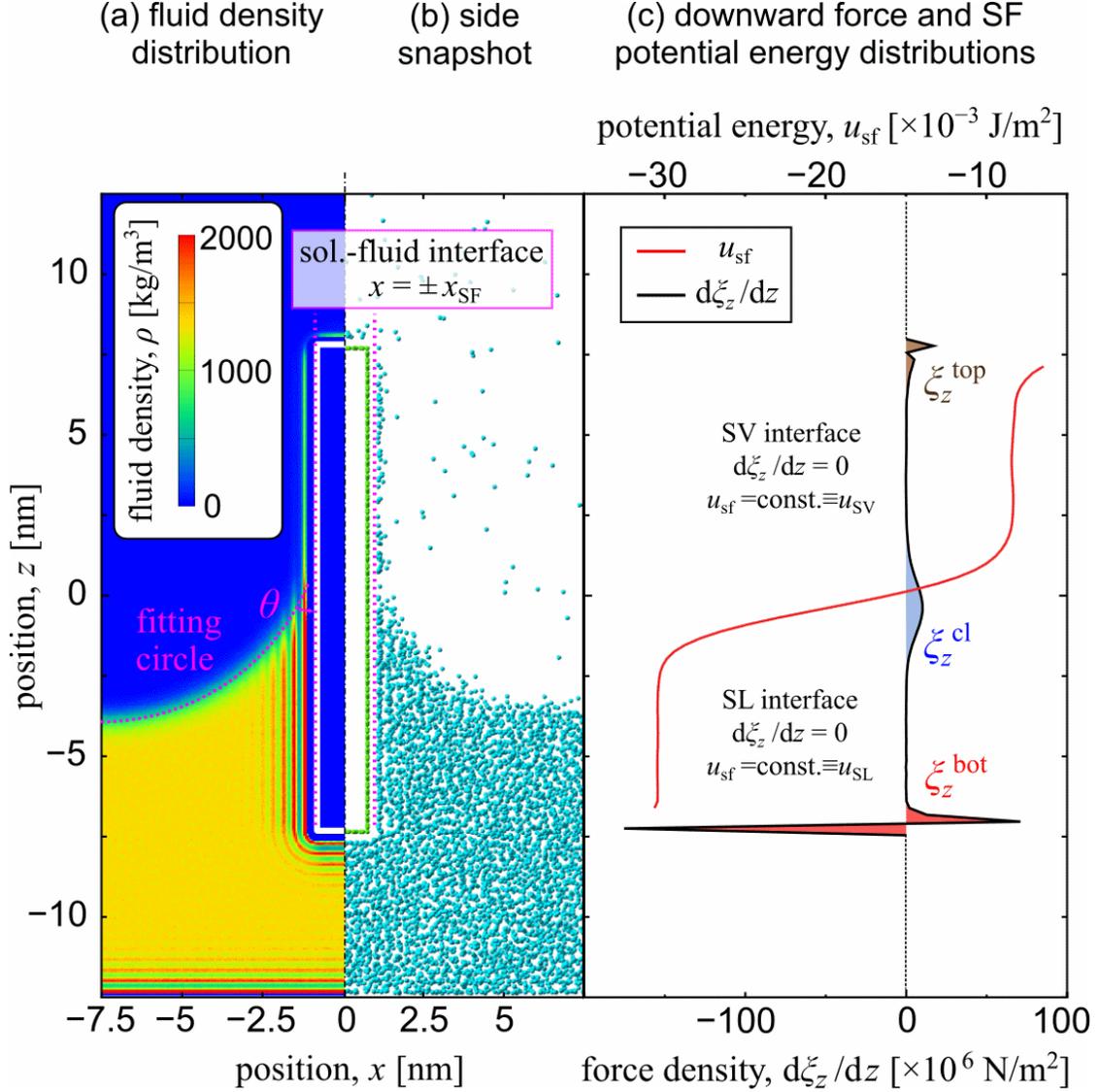}
 \caption{
(a) Distribution of the time-averaged fluid density, 
(b) half side snapshot, 
and (c) distributions of the time-averaged downward 
force density acting on the solid plate and solid-fluid (SF)
potential energy for the system with a
SF interaction parameter $\eta$ of 0.15.
}
\label{Fig:distribtution}
\end{figure}
We calculated the distribution of force exerted from the fluid 
on the solid particles by dividing the system into equal-sized  
bins normal to the $z$-direction, where the height of the bin 
$\delta z$ of 0.2115~nm was used considering the periodicity 
of the graphene structure.  
We defined the average force density $\dxizdz$
as the time-averaged total downward (in $-z$-direction) force 
from the fluid on the solid particles in each bin divided by 
$2l_y \delta z$, where $l_y$ is the system width in the $y$-direction. 
Except at the top and bottom of the solid plate, 
$\dxizdz$ corresponds to the total downward 
force from both sides divided by the sum of surface area of 
both sides, \ie the downward force per surface area. 
We also calculated the average SF potential 
energy per area $\usf$ as well, which was obtained 
by substituting the downward force by the SF potential 
energy.
\par
Figure~\ref{Fig:distribtution} shows the distribution of 
time-averaged fluid density  $\rho$ 
around the solid plate for the 
system with solid-fluid interaction parameter $\eta=0.15$
and a snapshot of the system. The time-averaged distributions of the downward force acting on the solid plate $\dxizdz$ and the SF potential energy $\usf$ are also displayed in the right panel. 
Multi-layered structures in the liquid, called the adsorption layers, were formed around the solid plate and the potential wall on the bottom, and liquid bulk with a homogeneous density is observed away from the potential wall, the solid plate and the LV interface.
%
\par
The downward force $\dxizdz$ on the solid 
plate in Fig.~\ref{Fig:distribtution}~(c)
was positive around the top as filled with brown, 
zero below the top up to around the CL, 
and had smoothly distributed positive values around the 
CL as filled with blue. 
As further 
going downward, it became zero again below around the CL, and 
showed sharp change from positive to negative values as filled 
with red. On the SV interface between the plate top and CL and on the SL interface between the CL and the plate bottom, the time-averaged downward force was zero.
Regarding the SF potential energy, $\usf$ was constant in the region where $\dxizdz=0$. 
This is because the  time-averaged fluid density in these regions 
was homogeneous in the $z$-direction, \ie 
$
\ptl \rho
/ \ptl z
=0
$
was satisfied within the range where the intermolecular force 
from the fluid on the solid particles effectively reaches,
and no surface-tangential force in the $z$-direction was 
exerted on the solid. 
This point will be described more in detail in Subsec.~\ref{subsec:analytical_xiz}.
Such two regions with zero downward force were formed for 
all systems in the present study, and thus, the total downward 
force as the integral of $\dxizdz$ can be clearly separated 
into three local parts, \ie $\xiztop$ 
around the top, $\xizcl$ around the contact line, 
and $\xizbot$ around the bottom. As indicated in 
Fig~\ref{Fig:distribtution}~(c), $\xiztop$ and $\xizcl$ are positive, 
\ie downward forces, and 
$\xizbot$ is negative, \ie an upward force.
Note that the distributions of $\dxizdz$ and $\usf$ 
around the top and bottom had less physical meaning because 
they included the top and bottom faces in the bin, and these parts for $\usf$ are not displayed in the figure.
However, the 
local integral of $\dxizdz$ indeed gave the physical information 
about the force around the top and bottom parts. Note also 
that $\xi_{z}$ has the same dimension as the surface 
tension of force per length.
\par
The LV interface had a uniform curvature away from 
the solid plate to minimize LV interface area as one of 
the principal properties of surface tension. 
Considering the symmetry of the system, the hemi-cylindrical LV interface with 
a uniform curvature is symmetrical  
between the solid plates over the periodic boundary in the 
$x$-direction.
Regarding SF interface position $\xsf$, which was different from the wall surface position $x_s$, we defined it at the limit that the fluid could reach. With this definition, Young's equation holds for quasi-2D droplets on a smooth and flat solid surface, as shown in our previous study.~\cite{Yamaguchi2019}
The $\xsf$ value was determined as $\xsf = 1.15$~nm from the 
density distribution, whereas the curvature radius $R$ was determined
through the least-squares fitting of a circle on the density contour 
of $\rho=$400~kg/m$^{3}$ at the LV interface excluding the region 
in the adsorption layers near the solid 
surface.~\cite{Nishida2014,Yamaguchi2019,Kusudo2019}
We defined the apparent contact angle $\theta$ by the angle at $x=\xsf$ between the SF interface and the least-squares fit of the LV interface having a curvature $\chi\equiv \pm 1/R$, with $R$ being the curvature radius. Note that the 
sign $\pm$ corresponds to the downward or upward convex LV-interfaces, respectively. 
The relation between the SF interaction coefficient $\eta$ and 
cosine of the contact angle $\cos \theta$ is shown in 
Appendix~\ref{sec:appendix_eta_costheta}, and the following results 
are shown based on $\cos \theta$ instead of $\eta$.
\begin{figure}
\centering
\includegraphics[width=0.6\linewidth]{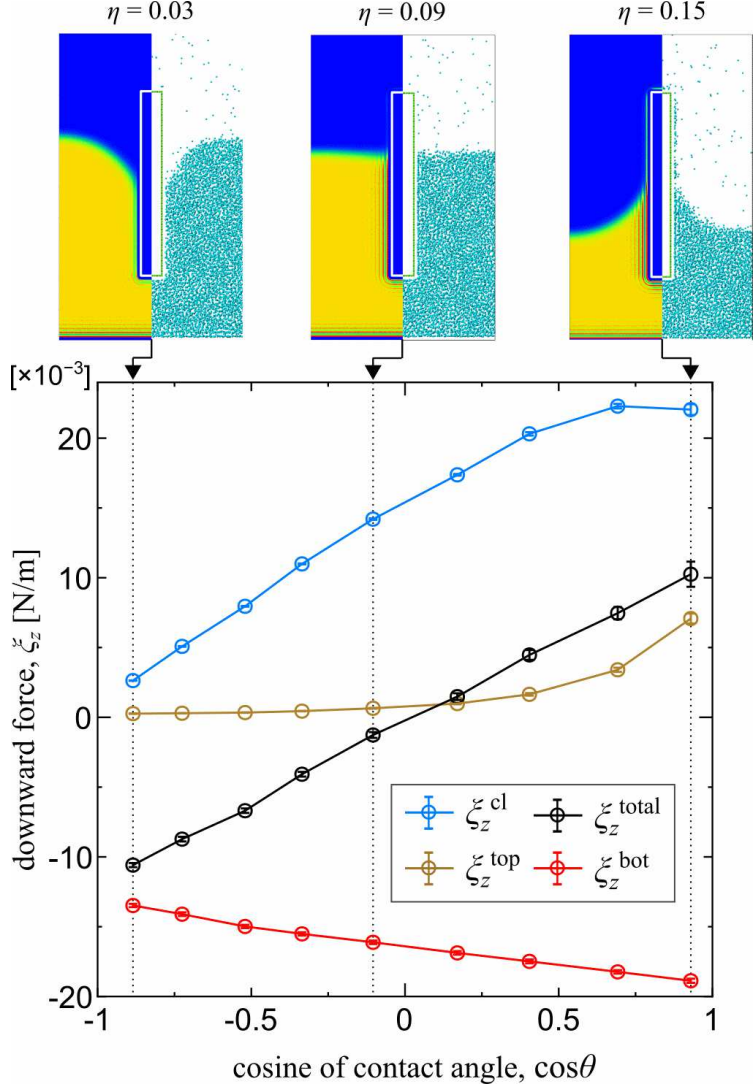}
\caption{MD results of the local downward 
forces exerted around the top, the contact line and the bottom of the solid plate and their sum as a function of 
the cosine of the contact angle. 
Corresponding half-snapshots and density distributions for three 
cases are also displayed on the top.
\label{Fig:local-forces}}
\end{figure}
\par
Figure~\ref{Fig:local-forces} shows the above-defined 
local downward forces $\xiztop$, $\xizcl$ and $\xizbot$ and 
their sum 
$\xiztot \equiv \xiztop + \xizcl + \xizbot$
on the cosine of the contact angle $\cos \theta$ obtained 
by MD simulations.
Corresponding half-snapshots and density distributions 
are also displayed on the top.
%
%
Regarding the force around the top $\xiztop$, 
it was almost zero except for cases with small contact angle.
This is obvious because almost no vapor particles were adsorbed on the 
top of the solid plate for non-wetting cases as seen in the top panel for $\eta = 0.03$.
However, in the case of large $\cos \theta$, $\xiztop$ had non-negligible positive value, \ie downward force comparable to $\xiztot$, because an adsorption layer was also formed at 
the SV interface as seen in the top panel for $\eta = 0.15$.
In terms of the force around the contact line $\xizcl$, it was positive even with negative $\cos \theta$ value, meaning
that the solid particle around the CL was always subject to a downward force from the fluid. 
%
On the contrary to $\xiztop$ and $\xizcl$, which were both positive, 
$\xizbot$ was negative and its magnitude increased as $\cos \theta$ increased, meaning that upward force to expel the bottom side was exerted from the liquid, and that the upward force was larger for larger SL interaction $\eta$.
Finally, the sum of the above three $\xiztot$ seems to be
proportional to $\cos\theta$. We will show later that it 
actually deviates from a simple 
Wilhelmy relation~\eqref{eq:Wilhelmy}.
\subsection{Analytical expressions of the forces on the solid}
\label{subsec:analytical_xiz}
\begin{figure}
\centering
\includegraphics[width=0.6\linewidth]{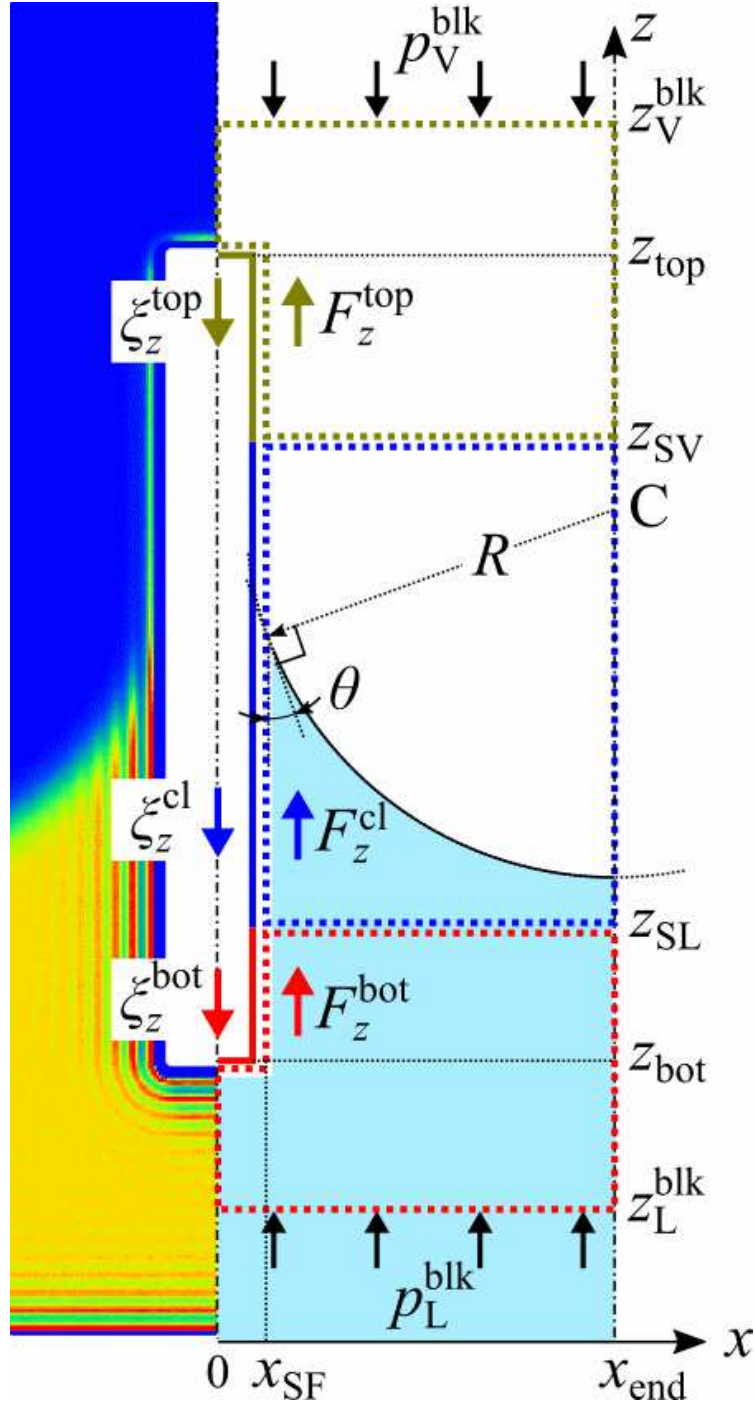}
\caption{
Top, contact-line (middle), and bottom parts of the solid plate 
subject to downward forces $\xiztop$, $\xizcl$ and $\xizbot$ from the 
fluid, respectively, and the control volumes (CVs) surrounding the 
fluid particles in contact with these solid parts subject to 
upward force $\Fztop$, $\Fzcl$ and $\Fzbot$ from the solid.
\label{Fig:controlvolume}
}
\end{figure}
\subsubsection{Definition of the solid-fluid forces}
In order to elucidate the origin of the forces exerted on the solid, 
we examined the details of the forces $\xiztop$, $\xizcl$ and $\xizbot$ 
from the fluid as well as the force balance on the control volumes (CVs)
surrounding the fluid around the solid plate with taking the stress 
distribution in the fluid into account as in our previous 
study.~\cite{Yamaguchi2019,Kusudo2019} 
We supposed three CVs surrounding 
the fluid around the solid plate as shown with dotted lines in 
Fig.~\ref{Fig:controlvolume}: a CV on the top in 
dark-yellow dotted line, one around the CL in blue dotted line,
and one on the bottom in red dotted line. All the CVs have 
their right face at the boundary of the system in the $x$-direction
at $x=x_\mathrm{end}$ at which symmetry of the physical values 
is satisfied, and the faces in contact with the solid is set 
at the limit that the fluid could reach. The remaining left 
sides of the top and bottom CVs are set in the center of the system
where the symmetry condition is satisfied. 
The $z$-normal faces are set respectively 
at $z=\zvblk$, $\zsv$, $\zsl$ 
and $\zlblk$, where $\zvblk$ 
and $\zlblk$ are at the vapor and liquid bulk 
heights, whereas $\zsv$ and $\zsl$ are set 
at the heights of SV and SL interfaces, respectively as shown in 
Fig.~\ref{Fig:local-forces} at which 
$\dxizdz = 0$ is satisfied. These heights can be set 
rather arbitrary as long as the above conditions are satisfied.
We define the forces from solid to liquid 
by $\Fztop$, $\Fzcl$ and $\Fzbot$ on the top, 
middle and bottom CVs, respectively. 
In addition, we also categorize the right-half of the solid plate 
into top, middle and bottom parts shown with dark-yellow, blue, 
and red solid lines, respectively with $\zsv$ and $\zsl$ as 
the boundaries as shown in Fig.~\ref{Fig:controlvolume}.
where forces $\xiztop$, 
$\xizcl$ and $\xizbot$ in the $z$-direction are exerted from the 
fluid, respectively. 
Specifically note that 
$\xizcl \neq \Fzcl$,  
$\xizbot \neq \Fzbot$ and 
$\xiztop \neq \Fztop$,  
because, for instance, $\Fzcl$ also includes the 
forces from the top and bottom parts of the solid, whereas
$\xizcl$ includes the forces from the top and bottom CVs.
In other words, the force between the middle solid part and 
middle fluid CV is in action-reaction relation, 
but $\Fzcl$ and $\xizcl$ include different extra forces 
above. This will be described more in detail in the following.
\par
\begin{figure}
\centering
\includegraphics[width=0.8\linewidth]{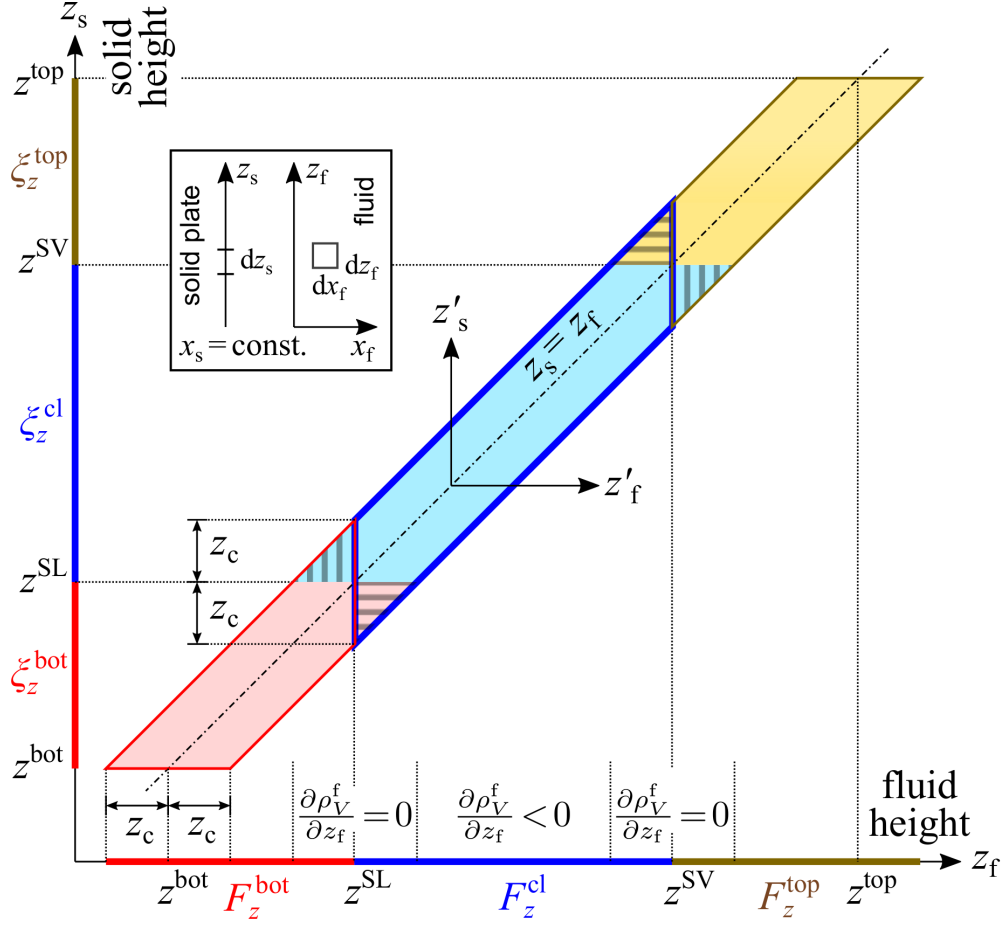}
\caption{Region for the double integral of the mean field regarding the 
interaction between solid plate and fluid at height $\zs$ and $\zf$, 
respectively. The geometrical relation is shown in the inset. 
Three height ranges of `top,' `cl,' and `bot' corresponding to those 
in Fig.~\ref{Fig:controlvolume} are depicted in color. 
Cutoff distance $\zc$ for $|\zf-\zs|$ is set depending on the lateral
position $\xf - \xs$, and the solid-liquid interactions between 
height ranges are categorized as filled regions or as ones surrounded 
by solid lines.
\label{Fig:meanpot_integ}
}
\end{figure}
\subsubsection{Capillary force $\xizcl$ around the contact line
based on a mean-field approach
\label{subsubsec:meanfield}
}
We start from formulating the wall tangential force on the 
solid particles $\xizcl$ on the right face of the solid 
plate. Taking into account that the 
solid is supposed to be smooth for the fluid particles 
because the interparticle distance parameters $\sigma_\mathrm{ff}$ 
and $\sigma_\mathrm{sf}$ are sufficiently large compared 
to $r_\mathrm{ss}$ between solid particles, $\xizcl$ can 
be analytically modeled by assuming the mean fields of 
the fluid and solid.
The mean number density per volume 
$\rnf(\zf,\xf)\ (=\rho/m_\mathrm{f})$ of the fluid is 
given as a function of the two-dimensional
position $(\zf,\xf)$ of the fluid, whereas a constant mean number density 
per area $\rns$ of the solid is used considering the present system
with a solid plate of zero-thickness without volume; however,  the
following derivation can easily be extended for a system with a solid 
with a volume and density per volume in the range $x \leq \xs$ as long as the density is independent of $\zs$.
We start from the potential energy on a solid particle at position $(\xs,\ys,\zs)$
due to a fluid particle at $(\xf,\yf,\zf)$ given by Eq.~\eqref{eq:LJ}. 
We define
\begin{equation}
    \xpf = -\xds \equiv \xf-\xs, \quad
    \ypf = -\yds \equiv \yf-\ys, \quad
    \zpf = -\zds \equiv \zf-\zs
    \label{eq:def_relpos}
\end{equation}  
in the following. 
Assuming that the fluid particles are homogeneously 
distributed in the $y$-direction with a number density $\rnf(\zf,\xf)$ per volume, the mean potential field  
from an infinitesimal volume segment of 
$\mathrm{d}\zf \times \mathrm{d}\xf$ 
on the solid particle 
is defined by using 
$\rnf(\zf,\xf)$
and the mean local potential $\phi(\zpf, \xpf)$
as 
$\rnf(\zf,\xf)\dzf \dxf
\cdot
\phi(\zpf, \xpf)$, 
where $\phi(\zpf, \xpf)$ is given by
\begin{equation}
\phi(\zpf, \xpf) \equiv 
\int_{-\infty}^{\infty} \Phi_\mathrm{LJ}(r)\dypf
 \label{eq:def_localphi}
\end{equation}
with 
\begin{equation}
r = \sqrt{\xpf^{2} + \ypf^{2} + \zpf^{2}},\quad
\sigma = \sigma_\mathrm{sf},\quad 
\epsilon = \epsilon_\mathrm{sf}.
\end{equation}
This schematic is shown in the inset of Fig.~\ref{Fig:meanpot_integ}.
Then, the local tangential force 
$f_{z}^\mathrm{s}(\zpf,\xpf)\dzf \dxf \dzs$
exerted on an infinitesimal 
solid area-segment of $\mathrm{d}\zs$ 
from the present fluid volume-segment is given by:
\begin{align}
\nonumber
f_{z}^\mathrm{s}(\zs,\zf,\xf)
\mathrm{d}\zf \mathrm{d}\xf \mathrm{d}\zs
&=
-\frac{\partial}{\partial \zs} \left[\rnf(\zf,\xf)\phi(\zpf, \xpf) \right]
\mathrm{d}\zf \mathrm{d}\xf \cdot \rns\mathrm{d}\zs
\\
&=
-\rns \rnf(\zf,\xf)
\frac{\partial \phi(\zpf, \xpf)}{\partial \zs}
\mathrm{d}\zf \mathrm{d}\xf \mathrm{d}\zs,
\label{eq:forcesegment}
\end{align}
where 
\begin{equation}
f_{z}^\mathrm{s}(\zs,\zf,\xf) = 
-\rns \rnf(\zf,\xf)
\frac{\partial \phi(\zpf, \xpf)}{\partial \zs}
\label{eq:forcedensity}
\end{equation}
denotes the tangential force density on the solid.
Note that $\mathrm{d}\xf$ and $\mathrm{d}\xpf$ are identical
because $\xs$ is a constant.
\par
Since $\Phi_\mathrm{LJ}(r)$ is truncated at the cutoff distance $\rc$ in the present case, 
\begin{gather}
\phi \left(\zpf, \xpf\right) 
=0,\quad
\frac{\ptl \phi\left(\zpf, \xpf\right)}{\ptl \zs} 
= 0
\label{eq:phi_limit}
\\
\mathrm{for}
\quad
|\zpf| \geq \sqrt{\rc^{2} - \xpf^{2}} \equiv
\zc(\xpf)
\quad \mathrm{or} \quad 
\xpf \geq \rc
\nonumber
\end{gather}
holds,
where $\zc(\xpf)$ as a function of $\xpf$ denotes the cutoff 
with respect to $\zpf$. 
Indeed this cutoff is not critical as long as 
$\phi\left(\zpf, \xpf\right)$ quickly vanishes 
with the increase of $r$, but we 
continue the derivation including the cutoff for simplicity.
With the definition of $x_\mathrm{SF}$ as the limit that the fluid could reach, it follows that 
\begin{equation}
\rnf=0 
\quad \mathrm{for} \quad
\xf < \xsf.
\end{equation}
%
%
In addition, considering that $\phi(\zpf, \xpf)$ is 
an even function with respect to $\zpf$, \ie
\begin{equation}
\phi\left(\zpf, \xpf\right)
=
\phi(-\zpf, \xpf),
\label{eq:phi_even}
\end{equation}
it follows for the mean local potential $\phi$ that
\begin{equation}
\frac{\partial \phi(\zpf, \xpf) }{\partial \zs} 
=
-\frac{\partial \phi(-\zpf, \xpf) }{\partial \zs}, 
\label{eq:phi_der_oddfunc}
\end{equation}
and
\begin{equation}
\frac{\partial \phi(\zpf, \xpf) }{\partial \zs} 
=
-\frac{\partial \phi(\zpf, \xpf) }{\partial \zf},
\label{eq:exchange_zs_zf}
\end{equation}
where Eq.~\eqref{eq:def_relpos} is applied for the latter,
which corresponds to the action-reaction relation between 
solid and fluid particles under a simple two-body 
interaction, \ie 
\begin{equation}
f_{z}^\mathrm{f}(\zs,\zf,\xf) 
=
-f_{z}^\mathrm{s}(\zs,\zf,\xf)
=
-\rns \rnf(\zf,\xf)
\frac{\partial \phi(\zpf, \xpf)}{\partial \zf}
\label{eq:forcedensity_f}
\end{equation}
holds for the tangential force density on the  fluid $f_{z}^\mathrm{f}$.
\par
Based on these properties, we now derive the analytical expression
of $\xizcl$ as the triple integral of the local tangential force 
$f_z^\mathrm{s}$ in Eq.~\eqref{eq:forcesegment}
around the CL, where the fluid density 
$\rnf$ decreases with the increase of $\zf$ within a certain range. 
Let this range be 
$\zsl + \zc \leq \zf \leq \zsv - \zc $ 
satisfying 
\begin{equation}
\frac{\ptl \rnf}{\ptl \zf} < 0 \quad (\zsl + \zc \leq \zf \leq \zsv - \zc ),
\label{eq:ptl_rnf_lt_0}
\end{equation}
and let $\rnf$ outside this range be given as a unique function of $\xf$ by
\begin{equation}
\rnf(\zf,\xf) =\left\{
\begin{array}{cc}
\rho_{V}^\mathrm{f(SL)}(\xf) & (\zsl - \zc < \zf < \zsl +  \zc)
\\
\rho_{V}^\mathrm{f(SV)}(\xf) & (\zsv - \zc < \zf < \zsv + \zc)
\end{array}
\right.
\label{eq:rnf_at_SL_SV}
\end{equation}
as shown in Fig.~\ref{Fig:meanpot_integ}.
Then, $\xizcl$ is expressed by
\begin{equation}
\xizcl\equiv
-\int_{\xsf}^{\xs+\rc} \left[
\int_{\zsl}^{\zsv}\left(
\int_{-\zc}^{\zc}
f_{z}^\mathrm{s}(\zs,\zpf,\xf) 
\mathrm{d}\zpf 
\right)\mathrm{d}\zs
\right] \mathrm{d}\xf
\label{eq:xizcl_tripleint}
\end{equation}
as the triple integral of the force density 
$f_{z}^\mathrm{s}$ in Eq.~\eqref{eq:forcedensity},
where the integration range of the double integral 
regarding $\zf$ and $\zs$ corresponds to the 
region filled with blue in Fig.~\ref{Fig:meanpot_integ}. 
\par
To obtain the double integral 
as the square brackets in Eq.~\eqref{eq:xizcl_tripleint} 
for the blue-filled region in Fig.~\ref{Fig:meanpot_integ}, 
we calculate at first that in the region surrounded by the solid-blue
line, add those in the vertically-hatched regions, and subtract
those in the horizontally-hatched regions.
Note that 
$\rnf(\zf,\xf) = \rho_{V}^\mathrm{f(SL)}(\xf)$ and 
$\rnf(\zf,\xf) = \rho_{V}^\mathrm{f(SV)}(\xf)$
are assumed for the hatched regions in the bottom-left 
and in the top-right regions, respectively
based on Eq.~\eqref{eq:rnf_at_SL_SV}.
%
The double integral for the region surrounded by 
the solid-blue line is
\begin{align}
\int_{\zsl}^{\zsv}\left(
\int_{-\zc}^{\zc}
f_{z}^\mathrm{s}\mathrm{d}\zds \right)\mathrm{d}\zf
&=
-\rns \int_{\zsl}^{\zsv}\!\!\!\!
\rnf(\zf,\xf)
\left(
\int_{-\zc}^{\zc}
\frac{\partial \phi(\zpf, \xpf)}{\partial \zs}\mathrm{d}\zds
 \right)\mathrm{d}\zf 
\nonumber
\\
&= 0,
\label{eq:dint_blue_fill}
\end{align}
by using Eq.~\eqref{eq:phi_even}.
%
Indeed, from Eq.~\eqref{eq:forcedensity_f}, the reaction 
force $-\Fzcl$ from solid on the fluid  around the CL in 
the blue-dotted line in Fig.~\ref{Fig:controlvolume} 
is obtained by further integrating 
Eq.~\eqref{eq:dint_blue_fill} 
with respect to $\xf$,
\ie 
\begin{align}
\int_{\xsf}^{\xs+\rc} \left[
\int_{\zsl}^{\zsv}\left(
\int_{-\zc}^{\zc}
f_{z}^\mathrm{s}\mathrm{d}\zds \right)\mathrm{d}\zf
\right] \mathrm{d}\xf
&= 
-\int_{\xsf}^{\xs+\rc} \left[
\int_{\zsl}^{\zsv}\left(
\int_{-\zc}^{\zc}
f_{z}^\mathrm{f}\mathrm{d}\zds \right)\mathrm{d}\zf
\right] \mathrm{d}\xf
\nonumber
\\
&=
-\Fzcl
\nonumber
\\
&=
0.
\label{eq:Fzcl=0}
\end{align}
The final equality means that no tangential force acts 
on the fluid there as mentioned  in our previous study.~\cite{Yamaguchi2019}
\par
Regarding the bottom-left vertically-hatched region in 
Fig.~\ref{Fig:meanpot_integ}, the double integral is
\begin{align}
\int_{-\zc}^{0}
\left(
\int_{-\zpf}^{\zc}
f_{z}^\mathrm{s}\mathrm{d}\zds \right)\mathrm{d}\zpf
&=
-\rns\rho_{V}^\mathrm{f(SL)}(\xf) \int_{-\zc}^{0}
\left(
\int_{-\zpf}^{\zc}
\frac{\partial \phi(\zpf, \xpf)}{\partial \zs}\mathrm{d}\zds
 \right)\mathrm{d}\zpf 
\nonumber    
\\ &=
\rns\rho_{V}^\mathrm{f(SL)}(\xf) \int_{-\zc}^{0}
\phi(\zpf, \xpf)\mathrm{d}\zpf ,
\label{eq:dint_bl_vert}
\end{align}
where $\phi(\zc, \xpf)=0$ and Eq.~\eqref{eq:phi_der_oddfunc} 
is used for the 2nd equality. This region physically 
corresponds to the interaction between blue solid part
and fluid in the red-dotted part in Fig.~\ref{Fig:controlvolume}.
For the bottom-left horizontally-hatched region in Fig.~\ref{Fig:meanpot_integ}, 
it follows that 
\begin{align}
\int_{0}^{\zc}\left(
\int_{-\zc}^{-\zpf}
f_{z}^\mathrm{s} \mathrm{d}\zds\right) \mathrm{d}\zf
&=
-\rns\rho_{V}^\mathrm{f(SL)}(\xf) \int_{0}^{\zc}
\left(
\int_{-\zc}^{-\zpf}
\frac{\partial \phi(\zpf, \xpf)}{\partial \zs}\mathrm{d}\zds
 \right)\mathrm{d}\zpf
\nonumber    
\\ &=
-\rns\rho_{V}^\mathrm{f(SL)}(\xf) \int_{0}^{\zc}
\phi(\zpf, \xpf)\mathrm{d}\zpf.
\label{eq:dint_bl_horiz}
\end{align}
%
This region corresponds to the interaction between 
red solid part and fluid in the blue-dotted part in 
Fig.~\ref{Fig:controlvolume}.
Hence, the net force due to the double integral in the 
bottom-left hatched regions in Eqs.~\eqref{eq:dint_bl_vert} 
and \eqref{eq:dint_bl_horiz} with also integrating in the 
$\xf$-direction, which we define by $\usl$, 
results in
\begin{equation}
\usl\equiv
    \rns\int_{0}^{\rc} 
    \left(\rho_{V}^\mathrm{f(SL)}(\xpf) 
    \int_{-\zc(\xpf)}^{\zc(\xpf)}
\phi(\zpf, \xpf)\mathrm{d}\zpf\right)\mathrm{d}\xpf.
\label{eq:def_usl}
\end{equation}
As a physical meaning, $\usl$ represents the SL
potential energy density \ie potential energy per 
SL-interfacial area at the SL interface away from 
the CL and from the bottom of the solid plate.
\par
Regarding the top-right hatched regions, the net force 
results in $-\usv$ with the SV potential energy
density area given by
\begin{equation}
\usv\equiv
    \rns\int_{0}^{\rc} 
    \left(\rho_{V}^\mathrm{f(SV)}(\xpf) 
    \int_{-\zc(\xpf)}^{\zc(\xpf)}
\phi(\zpf, \xpf)\mathrm{d}\zpf\right)\mathrm{d}\xpf,
\label{eq:def_usv}
\end{equation}
which can be derived in a similar manner.
Thus, it follows for the force  $-\xizcl$ from 
the fluid on the solid around the CL that 
\begin{equation}
    -\xizcl = -\Fzcl + \usl - \usv,
\quad
    \xizcl = \Fzcl - \usl + \usv,
\label{eq:xizcl_Fzcl}
\end{equation}
therefore, by using $\Fzcl=0$ in Eq.~\eqref{eq:Fzcl=0},  
\begin{equation}
    \xizcl =  - \usl + \usv = (- \usl) - (-\usv)
\label{eq:xizcl_eq_potdif}
\end{equation}
is derived as the analytical expression of $\xizcl$, where 
the final expression is appended considering that the 
potential energy densities $\usl$ and $\usv$ are both negative.
\par
\begin{figure}
\centering
\includegraphics[width=0.8\linewidth]{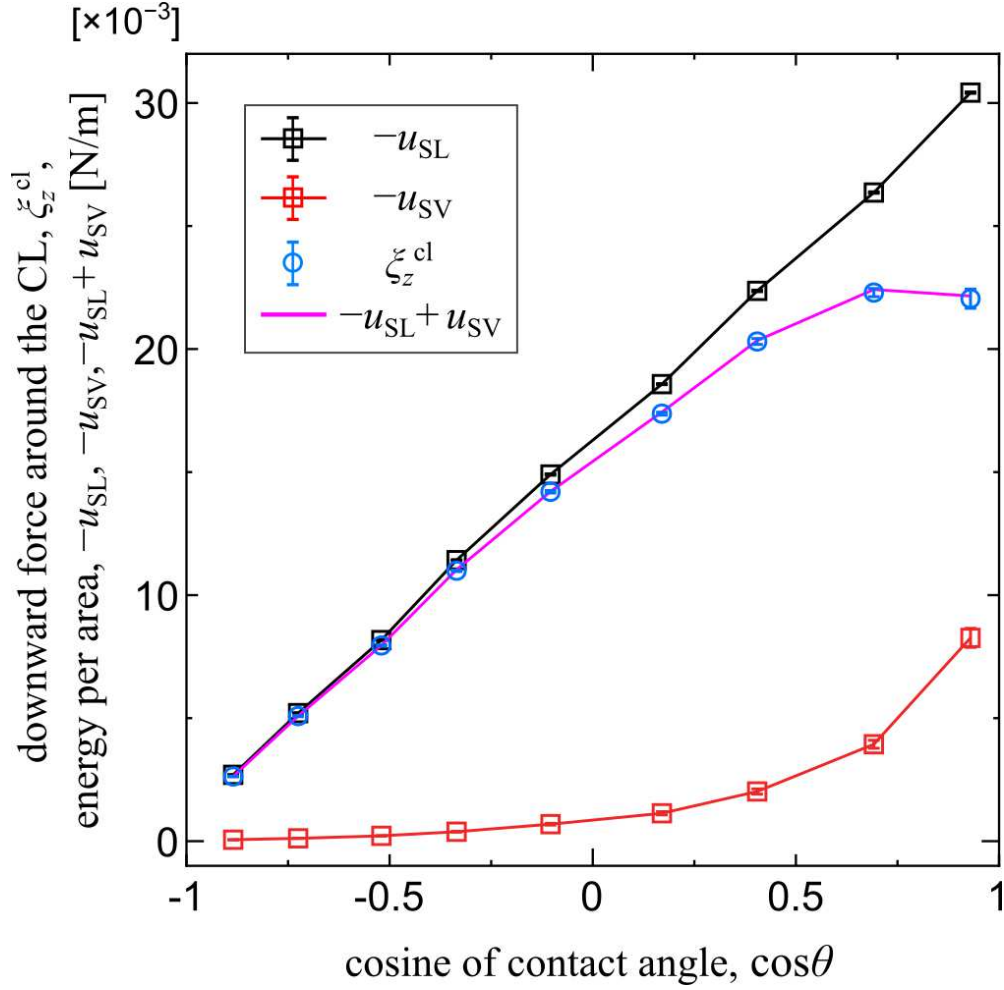}
\caption{
Dependence of the SL and SV potential density energy densities 
$\usl$ and $\usv$ as the potential energies per interfacial area on the cosine of the 
contact angle $\cos \theta$, 
and comparison between the force on the solid around the CL 
$\xizcl$ and difference of potential energy density
$- \usl + \usv$.
}
\label{Fig:usl_usv}
\end{figure}
Figure~\ref{Fig:usl_usv} shows the dependence of the 
SL and SV potential energy density $\usl$ 
and $\usv$, respectively as the potential energies 
per interfacial area, on the cosine of the 
contact angle $\cos \theta$, 
and comparison between the force on the solid around the CL 
$\xizcl$ and difference of potential energy density
$- \usl + \usv$.
Very good agreement between $\xizcl$ and $- \usl + \usv$ 
is observed within the whole range of the contact angle, 
and this indicates that Eq.~\ref{eq:xizcl_eq_potdif} is 
applicable for the present system with a flat and smooth 
surface. It is also qualitatively apparent from
Eq.~\eqref{eq:xizcl_eq_potdif}
that $\xizcl$ is positive regardless of the contact angle 
because the SF potential energy is smaller at the SL 
interface than at the SV interface.
It is also interesting to note that for the very wettable case 
with large $\cos \theta$, \ie large wettability parameter 
$\eta$, $\xizcl$ decreased with the increase of $\cos \theta$. 
This can be explained as follows: the change of $-\usv$ and 
$-\usl$ are both due to the change of $\eta$ and the fluid 
density especially in the first adsorption layer, while the 
density change of the SL adsorption layer due to $\eta$ is 
rather small. Thus, for higher $\eta$ value, the effect of 
density increase of the SV adsorption layer on $-\usv$ 
upon the increase of $\eta$ overcomes the increase of 
$-\usl$.
\subsubsection{Total force $\xiztot$ and local forces $\xizbot$ and $\xiztop$ on the bottom and the top}
Before proceeding to the analytical expression of 
$\xizbot$ and $\xiztop$, we derive their relations with 
$\Fzbot$ and $\Fztop$. Through the comparison between 
the regions of double integration for $\xizbot$ and $\Fzbot$
with respect to $\zf$ and $\zs$ in Fig.~\ref{Fig:meanpot_integ}, 
\ie the red-filled region and one surrounded by solid-red line,
it is clear that the difference between $\xizbot$ and $\Fzbot$ 
corresponds to the integral of hatched regions around $\zsl$ in the 
bottom-left. Thus, it follows that 
\begin{equation}
    \xizbot = \Fzbot + \usl
    \label{eq:xizbot_Fzbot}
\end{equation}
and 
\begin{equation}
    \xiztop = \Fztop - \usv.
    \label{eq:xiztop_Fztop}
\end{equation}
Note that the sum of Eqs.~\eqref{eq:xizcl_eq_potdif},  
\eqref{eq:xizbot_Fzbot} and \eqref{eq:xiztop_Fztop}
satisfies 
\begin{equation}
\xiztot = F_{z}^\mathrm{top} + F_{z}^\mathrm{cl} + F_{z}^\mathrm{bot}.
\label{eq:xiztot=Fztot}
\end{equation}
%
Considering that feature, we examine 
the total force $\xiztot$ and 
local ones $\xizbot$ and $\xiztop$ on the bottom and the top.
We consider the distribution of the two-dimensional fluid 
stress tensor $\bm{\tau}$ averaged in the $y$-direction by the 
method of plane (MoP)~\cite{Thompson1984, Yaguchi2010} based 
on the expression by \citet{Irving1950} (IK), 
with which exact force balance is satisfied for an arbitrary control 
volume bounded by a closed surface. 
The stress tensor component $\tau_{\alpha\beta}(x,z)$ denotes
the stress in $\beta$-direction exerted on an infinitesimal 
surface element with an outward normal in $\alpha$-direction 
at position $(x,z)$. 
In the formulation of the MoP based on the IK-expression,  
$\tau_{\alpha\beta}(x,z)$ consists of the time-average of the kinetic 
and inter-molecular interaction contributions due to the molecular 
motion passing through the surface element and the intermolecular 
force crossing the surface element, respectively. For a single 
mono-atomic fluid component whose constituent particles interact 
through a pair potential as in the present study, all force line 
segments between two fluid particles, which cross the surface 
element, are included in the second. 
Note that technically for the MoP, the SF interaction can also be 
included in the inter-molecular force contribution, but only the 
FF interaction as the internal force is taken into account as
the stress, and SF contribution is considered as an external force in this study.~\cite{Nijmeijer1990_simul,Schofield1982,Rowlinson1993,Yamaguchi2019, Kusudo2019} 
With this setting, the stress is zero at the SF boundary 
for all CVs because no fluid particle exists beyond the 
boundary to contribute to the stress component as the kinetic 
nor at inter-molecular interaction contribution. Hence, 
the force balance on each CV containing only fluid is 
satisfied with the sum of the stress surface integral
and external force from the solid. 
The force balance on the red-dotted CV
in Fig.~\ref{Fig:controlvolume} in the $z$-direction is 
expressed by
\begin{equation}
    - \int_{0}^{\xend}\tau_{zz}(x,\zlblk) \drom x
    + \int_{\xsf}^{\xend}\tau_{zz}(x,\zsl) \drom x
    + \Fzbot = 0,
\label{eq:forcebalance_botCV}
\end{equation}
with the stress contributions from the bottom and top
and external force in the RHS, respectively, by taking 
into account that $\tau_{xz}=0$ on the 
$x$-normal faces at $x=0$ and $x=\xend$ due to the 
symmetry, and also that the stress at the SF interface 
is zero. Similarly, the force balance on the blue-dotted CV
and dark-yellow-dotted CV in Fig.~\ref{Fig:controlvolume}
in the $z$-direction are 
expressed by
\begin{equation}
    - \int_{\xsf}^{\xend}\tau_{zz}(x,\zsl) \drom x
    + \int_{\xsf}^{\xend}\tau_{zz}(x,\zsv) \drom x
    + \Fzcl = 0,
\label{eq:forcebalance_midCV}
\end{equation}
and 
\begin{equation}
    - \int_{\xsf}^{\xend}\tau_{zz}(x,\zsv) \drom x
    + \int_{0}^{\xend}\tau_{zz}(x,\zvblk) \drom x
    + \Fztop = 0,
\label{eq:forcebalance_topCV}
\end{equation}
respectively.
\par
By taking the sum of Eqs.~\eqref{eq:forcebalance_botCV}, 
\eqref{eq:forcebalance_midCV} and \eqref{eq:forcebalance_topCV}, 
and inserting Eq.~\eqref{eq:xiztot=Fztot}, it follows for $\xiztot$
that
\begin{equation}
    \xiztot
    = \int_{0}^{\xend}\tau_{zz}(x,\zlblk) \drom x
    - \int_{0}^{\xend}\tau_{zz}(x,\zvblk) \drom x
    \label{eq:xiztbot_stress}
\end{equation}
Since the bottom face of the red-dotted CV and top face 
of the dark-yellow-dotted CV in Fig.~\ref{Fig:controlvolume}
are respectively set in the liquid and vapor bulk regions under an 
isotropic static pressure $\plblk$, and $\pvblk$ given by
\begin{equation}
\plblk = -\tau_{xx}(x,\zlblk) = -\tau_{zz}(x,\zlblk),
\label{eq:plblk_tauzlblk}
\end{equation}
and 
\begin{equation}
\pvblk = -\tau_{xx}(x,\zvblk) = -\tau_{zz}(x,\zvblk),
\label{eq:pvblk_tauzvblk}
\end{equation}
the 1st and 2nd terms in the RHS of Eq.~\eqref{eq:xiztbot_stress} 
write
\begin{equation}
\int_{0}^{\xend}\tau_{zz}(x,\zlblk) \drom x
= -\int_{0}^{\xend} \plblk \drom x
= -\plblk \xend,
\label{eq:stressint_bot}
\end{equation}
and
\begin{equation}
\int_{0}^{\xend}\tau_{zz}(x,\zvblk) \drom x
= -\int_{0}^{\xend} \pvblk \drom x
= -\pvblk \xend.
\label{eq:stressint_top}
\end{equation}
Thus, Eq.~\eqref{eq:xiztbot_stress} results in a simple 
analytical expression of
\begin{equation}
    \xiztot
    = (\pvblk-\plblk) \xend.
    \label{eq:xiztot_laplacepressure}
\end{equation}
Furthermore, by applying the geometric relation
\begin{equation}
\sin\left(\theta -\frac{\pi}{2}\right) =
\cos \theta = 
\chi\left( \xend - \xsf \right)
\label{eq:geom_curv}
\end{equation}
with $\chi$ being the LV interface curvature and the Young-Laplace equation for the pressure 
difference in Eq.~\eqref{eq:xiztot_laplacepressure}:

\begin{equation}
\pvblk - \plblk = \glv\chi
= \frac{\glv\cos \theta}{\xend - \xsf},
\label{eq:Young-Laplace}
\end{equation}
which hold irrespective of whether the LV-interface 
is convex downward or upward, 
it follows for Eq.~\eqref{eq:xiztot_laplacepressure}
as another analytical expression of $\xiztot$ that
\begin{equation}
\xiztot
=
\frac{\xend}{\xend-\xsf}
\glv \cos \theta,
\label{eq:nano-Wilhelmy}
\end{equation}
which includes the correction to Eq.~\eqref{eq:Wilhelmy}
considering the effect of the Laplace pressure due to 
the finite system configuration with the periodic 
boundary condition.
Note also that from Eq.~\eqref{eq:nano-Wilhelmy}, 
by giving $\xend$ and $\xsf$, it is possible to 
estimate $\glv$ from the relation between 
$\xiztot$ and $\cos \theta$.
\par
\begin{figure}
\centering
\includegraphics[width=0.8\linewidth]{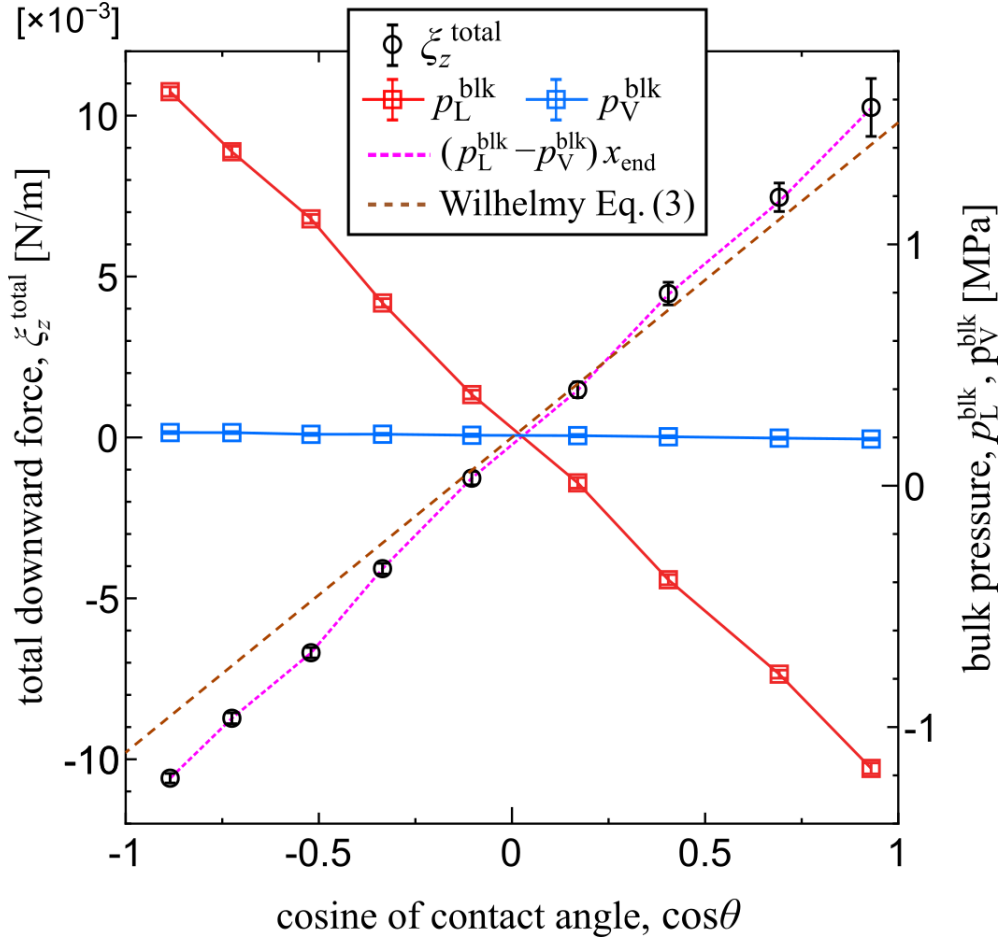}
\caption{Comparison of the total downward force $\xiztot$
on the solid plate directly obtained from MD with the 
analytical expression $(\pvblk-\plblk) \xend$ in 
Eq.~\eqref{eq:xiztot_laplacepressure} using the pressures 
$\plblk$ and $\pvblk$ measured on the bottom and top 
boundaries.
The Wilhelmy equation~\eqref{eq:Wilhelmy} using 
$\glv=9.79\times 10^{-3}$~N/m evaluated by the 
Young-Laplace equation~\eqref{eq:Young-Laplace} is also shown.
\label{Fig:comparison-wil-model}
}
\end{figure}
Figure~\ref{Fig:comparison-wil-model} shows
the comparison of the total downward force $\xiztot$
on the solid plate directly obtained from MD with the 
analytical expression $(\pvblk-\plblk) \xend$ in 
Eq.~\eqref{eq:xiztot_laplacepressure} using the pressures 
$\plblk$ and $\pvblk$ measured on the bottom and top 
boundaries as the force exerted from the fluid on the 
potential walls per area. Clearly $\xiztot$ and
$(\pvblk-\plblk) \xend$ agree very well, and this is
because Eq.~\eqref{eq:xiztot_laplacepressure} is simply 
the force balance to be satisfied for equilibrium systems.
Regarding the pressure, $\pvblk$ is almost constant, which 
corresponds to the saturated vapor pressure at this 
temperature. In addition, a linear relation between 
$\plblk - \pvblk$ and $\cos \theta$ can be observed, and 
this indicates that the Young-Laplace 
equation~\eqref{eq:Young-Laplace} is applicable in the 
present scale. We evaluated $\glv$ from this relation with 
the least-squares fitting, and the resulting value 
was $\glv = 9.79 \pm 0.23 \times 10^{-3}$~N/m with 
$\xsf = 1.15$~nm and $\xend=7.5$~nm, which was indeed close to the value obtained by a standard 
mechanical process.~\cite{Surblys2014}
The standard Wilhelmy equation~\eqref{eq:Wilhelmy} using this value is also shown in Fig.~\ref{Fig:comparison-wil-model}, 
indicating that 
$\glv$ would be 
overestimated with this standard Wilhelmy equation~\eqref{eq:Wilhelmy} in a small measurement system like the present one.
\par
Finally, we derive the analytical expression of the local 
force $\xizbot$ and $\xiztop$. 
For the derivation of $\xizbot$, we apply the extended Bakker's equation 
for the SL relative interfacial tension~\cite{Yamaguchi2019,Kusudo2019}
\begin{equation}
\gsl - \gs0 
=
\int_{\xsf}^{\xend} 
\left[\tau_{zz}(x,\zsl)-\tau_\mathrm{L}^\mathrm{blk}\right] \drom x
\label{eq:bakker_SL}
\end{equation}
for the 2nd term in the LHS of Eq.~\eqref{eq:forcebalance_botCV}, 
where $\gsl - \gs0$ is the SL interfacial tension 
relative to the interfacial tension between solid and fluid 
with only repulsive interaction (denoted by ``0" to 
express the solid surface without adsorbed fluid particles). 
Then, it follows 
that
\begin{equation}
\int_{\xsf}^{\xend} 
\tau_{zz}(x,\zsl) \drom x
=
\gsl - \gs0 - (\xend-\xsf)\plblk.
\label{eq:stressint_SL}
\end{equation}
By inserting Eqs.~\eqref{eq:xizbot_Fzbot}, 
\eqref{eq:stressint_bot}  and 
\eqref{eq:stressint_SL} 
into Eq.~\eqref{eq:forcebalance_botCV}, 
the analytical expression of $\xizbot$ writes
\begin{align}
    \xizbot 
    &= -\plblk \xend
    - [\gsl - \gs0 - (\xend-\xsf)\plblk]
    + \usl
    \nonumber
    \\
    &=
    -\xsf \plblk - (\gsl - \gs0) + \usl.
    \label{eq:xizbot_final}
\end{align}
Similary, by applying the Extended Bakker's equation for the SV interfacial tension~\cite{Yamaguchi2019,Kusudo2019}
\begin{equation}
\gsv - \gs0 
=
\int_{\xsf}^{\xend} 
\left[\tau_{zz}(x,\zsv)-\tau_\mathrm{V}^\mathrm{blk}\right] \drom x
\label{eq:bakker_SV}
\end{equation}
to Eq.~\eqref{eq:forcebalance_topCV} with 
Eq.~\eqref{eq:xiztop_Fztop}, the analytical expression of $\xiztop$ writes
\begin{equation}
    \xiztop
    =
    \xsf\pvblk + (\gsv - \gs0) - \usv.
    \label{eq:xiztop_final}
\end{equation}
\par
To verify Eqs.~\eqref{eq:xizbot_final} and \eqref{eq:xiztop_final}, 
we compared the present results with $\xizbot$ and $\xiztop$ 
calculated using the corresponding SL and SV works of 
adhesion $\Wsl$ and $\Wsv$ obtained by the thermodynamics integration 
(TI) with the dry-surface scheme.~\cite{Leroy2015,Yamaguchi2019}
The calculation detail is shown in Appendix~\ref{sec:appendix_TI}.
By definition, the SL and SV interfacial tensions $\gsl$ and $\gsv$ 
are related to $\Wsl$ and $\Wsv$ by
\begin{equation}
   W_\mathrm{SL} 
   \equiv
   \gs0 + \gl0 - \gsl
   \approx
   \gs0 + \glv - \gsl
   \label{eq:W_sl}
\end{equation}
and 
\begin{equation}
   \Wsv
   \equiv
   \gs0 + \gv0 - \gsv
  \approx 
   \gs0 - \gsv,
\label{eq:W_sv}
\end{equation}
respectively, where 
the approximation 
%
$
    \gl0 \approx \glv
$
%
for the interfacial tension $\gl0$ 
between liquid and vacuum is used in Eq.~\eqref{eq:W_sl}, 
and $\gv0$ is set zero in the final approximation in Eq.~\eqref{eq:W_sv}. 
Note that $\gl0$ or $\glv$ is included in $\Wsl$.
From Eqs.~\eqref{eq:W_sl} and \eqref{eq:xizbot_final}, 
and from Eqs.~\eqref{eq:W_sv} and \eqref{eq:xiztop_final}, 
$\xizbot$ and $\xiztop$ are respectively rewritten by
\begin{equation}
\xizbot \approx \Wsl -\plblk \xsf - \glv + \usl,
\label{eq:xizbot_Wsl_for_fig8}
\end{equation}
and
\begin{equation}
\xiztop \approx \xsf\pvblk - \Wsv - \usv.
\label{eq:xiztop_Wsv_for_fig8}
\end{equation}
\par
\begin{figure}
\includegraphics[width=0.8\linewidth]{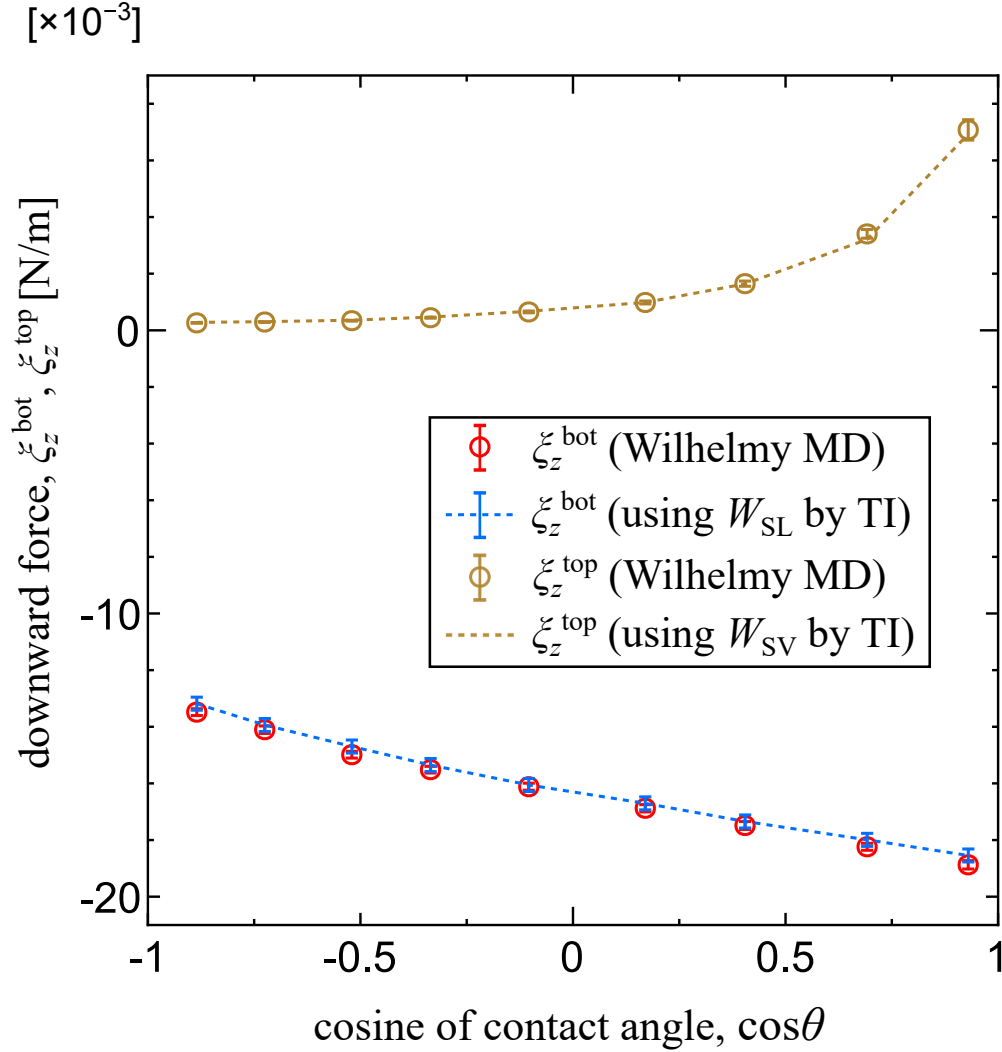}
\caption{Comparison of the downward forces $\xizbot$ and 
$\xiztop$ on the bottom and top of the solid plate directly obtained 
from MD with those evaluated using the works of adhesion $\Wsl$ and 
$\Wsv$ calculated by the thermodynamic integration (TI) using the 
dry-surface scheme shown in Appendix~\ref{sec:appendix_TI}. 
The error bar for $\xizbot$ using $\Wsl$ in blue comes from the evaluation 
of $\glv$ from $\plblk$ and $\pvblk$ in Fig.~\ref{Fig:comparison-wil-model}.
\label{Fig:fig08-bot-top-Wil-TI}
}
\end{figure}
Figure~\ref{Fig:fig08-bot-top-Wil-TI}
shows the comparison of $\xizbot$ and $\xiztop$ 
directly obtained from MD with those evaluated by 
Eqs.~\eqref{eq:xizbot_Wsl_for_fig8} and 
\eqref{eq:xiztop_Wsv_for_fig8} using the 
SL and SV works of adhesion $\Wsl$ and $\Wsv$, respectively 
obtained by the TI with the DS scheme shown in 
Appendix~\ref{sec:appendix_TI}.
Note that except $\Wsl$ and $\Wsv$, we used  
the values of $\plblk$, $\pvblk$, $\xend$, $\usl$, and $\usv$ 
directly obtained from the present Wilhelmy MD simulations 
as well as the $\glv$ value evaluated in 
Fig.~\ref{Fig:comparison-wil-model}. The error bars for $\xizbot$ using $\Wsl$ in blue mainly came from the error upon evaluating $\glv$.  
Note also that the TI calculation in Appendix~\ref{sec:appendix_TI} 
for $\Wsl$ was carried out under a control pressure of 1~MPa 
whereas that for $\Wsv$ was considered to be under the saturated 
vapor pressure at the present temperature.
For both $\xizbot$ and $\xiztop$, the Wilhelmy MD and TI results 
agreed well, and this indicates the validity of the
present analytical expression.
%
%
\subsection{Discussion}
We list the key issues for the further application of the present expression
in the following. 
First, 
Eqs.~\eqref{eq:forcebalance_botCV}, \eqref{eq:forcebalance_midCV} and \eqref{eq:forcebalance_topCV} 
are about the force balance and should be satisfied in equilibrium systems without any restrictions. In addition, Eqs.~\eqref{eq:xizcl_Fzcl}, \eqref{eq:xizbot_Fzbot}
and \eqref{eq:xiztop_Fztop} are about the relation between the solid-fluid and fluid-solid forces and should hold as long as 
the solid plate can be decomposed into the three parts without the interface overlapping.
At both SL and SV interfaces, which are between the CL and the plate bottom and between CL and the plate top respectively, a quasi-one-dimensional density 
distribution with 
$\ptl \rho/\ptl z=0$ can be assumed and one can apply the mean-field approach described in Sec.~\ref{subsubsec:meanfield}. 
Furthermore, Eqs.~\eqref{eq:bakker_SL}
and \eqref{eq:bakker_SV} are  Extended Bakker's 
equations~\cite{Yamaguchi2019} for the SL and SV interfacial tensions.
Hence, our analytical expressions with these equations 
are constructed by a purely mechanical approach, 
and are exact, as observed in the comparison 
in Figs.~\ref{Fig:usl_usv} and \ref{Fig:comparison-wil-model}.
\par
Another issue is about the relation
between Young's equation~\eqref{eq:Young} and the Wilhelmy equation~\eqref{eq:xiztot_laplacepressure} formulated with the Laplace pressure. 
Indeed, Eq.~\eqref{eq:xiztot_laplacepressure} holds irrespective of whether 
the CL is pinned or not because
this relation means a simple 
equilibrium force balance. 
In the present case, $\Fzcl=0$ in Eq.~\eqref{eq:Fzcl=0}
is satisfied because the solid surface is flat and smooth, 
and Young's equation holds. This can easily be proved 
considering the force balance in 
Eq.~\eqref{eq:forcebalance_midCV} about the middle CV. 
In cases with $\Fzcl\neq 0$ because of the pinning
force exerted on the fluid from the solid
around the CL, \textit{e.g.,} due to the boundary of 
wettability parallel to the CL in our 
previous research,~\cite{Kusudo2019} 
Young's equation should be rewritten including the 
pinning force. 
Even if such wettability boundary would 
be included in the present system, 
Eq.~\eqref{eq:xiztot_laplacepressure} 
would still be satisfied. In practice, 
such pinning force denoted by $\zpin$ 
in Ref.~\citenum{Kusudo2019} as the downward 
force from the solid on the fluid around the
CL corresponds to $-\Fzcl$ here, and this 
can be extracted by Eq.~\eqref{eq:xizcl_Fzcl} as
\begin{equation}
    -\zpin = \Fzcl = \xizcl + \usl - \usv.
\end{equation}
\par
Considering the above discussion, we summarize the 
procedure to extract the wetting properties.
In a single Wilhelmy MD simulation, 
we can calculate
\begin{enumerate}
\item
Force $\xiztop$, $\xizcl$ and $\xizbot$ on three 
parts of the solid
from the force-density distribution $\dxizdz$ in the 
surface-tangential direction,
\item
SF potential energy densities $\usl$ and $\usv$ on 
solid per area at SL and SV interfaces, respectively
from the distribution of the potential energy 
density $\usf$,
\item
Bulk pressures $\pvblk$ and $\plblk$ 
measured on the top and bottom of the system, and 
\item
Contact angle $\theta$ from the density distribution.
\end{enumerate}
%
From these quantities the following physical properties can be obtained:
\begin{enumerate}
\renewcommand{\labelenumi}{\alph{enumi}.}
\item
SL relative interfacial tension $\gsl - \gs0$ 
from $\xizbot$, $\usl$, $\xsf$ and $\plblk$ 
using Eq.~\eqref{eq:xizbot_final},
\item
SV relative interfacial tension $\gsv - \gs0$ from $\xiztop$,  $\usv$, $\xsf$ and $\pvblk$
using Eq.~\eqref{eq:xiztop_final},
\item 
LV interfacial tension $\glv$ from $\pvblk$, $\plblk$, $\xsf$, the system size $\xend$ and the contact angle $\theta$ using Eq.~\eqref{eq:Young-Laplace} , and
\item
Pinning force $\Fzcl$ from  Eq.~\eqref{eq:xizcl_Fzcl} to be added to Young's equation, which is zero in the case of flat and smooth
solid surface.
\end{enumerate}
Related to  the above procedure, it should also be noted 
that, surprisingly, the microscopic structure of the bottom 
face does not have a direct effect on the force $\xizbot$. 
This is similar to buoyancy given by the 3rd 
term of the RHS of 
Eq.~\eqref{eq:Wilhelmy_full}, which depends on the volume $V$
immersed into the liquid and is not directly related to the microscopic structure.
%
\par
Finally, we compare the present analytical expression of 
the contact line force $\xizcl$ with an 
existing model by \citet{Das2011}, which states
\begin{equation}
\xizcl = \gsv - \gsl + \glv =\glv (1+\cos\theta).
\label{eq:das_model}
\end{equation}
This model is derived based on the assumption that the densities 
of the liquid and vapor are constant at bulk values even close to the solid interface: the so-called sharp-kink approximation. 
This is similar to the interface of two different solids 
whose densities and structures do not change upon contact. 
Even under this assumption, the force $\xizcl$ on solid 
around the CL is expressed by Eq.~\eqref{eq:xizcl_eq_potdif} as the difference between the SL and SV potential energy densities $\usl$ and $\usv$ as well.~\cite{Das2011} The difference arises for the works 
of adhesion. Under the sharp-kink approximation, it is 
clear that the works of adhesion required to quasi-statically 
strip the liquid and vapor off the solid surface are equal 
to the difference of solid-fluid potential energies after 
and before the procedure, \ie
\begin{equation}
\Wsl = 0 -\usl = -\usl, \quad
\Wsv= 0 -\usv = -\usv
\quad\mbox{(under the sharp-kink approx.),}
\label{eq:wsl_usl_wsv_usv}
\end{equation} 
because the solid and fluid structures do not change upon
this procedure.
Then, it follows for Eq.~\eqref{eq:xizcl_eq_potdif} that
\begin{equation}
\xizcl = \Wsl - \Wsv 
\quad\mbox{(under the sharp-kink approx.),}
\end{equation}
which indeed results in Eq.~\eqref{eq:das_model} with 
Eqs~\eqref{eq:W_sl} and \eqref{eq:W_sv}.
However, the density around the solid surface 
is not constant as shown in the 
density distribution in Fig.~\ref{Fig:distribtution}, and
the difference of $\Wsl$ and $\Wsv$ is not directly 
related to the SL and SV potential energy densities 
$\usl$ and $\usv$ as in Eq.~\eqref{eq:wsl_usl_wsv_usv}.
In other words, the fluid can freely deform and can have 
inhomogeneous density in a field formed by the solid at 
the interface to minimize its free energy at equilibrium, 
and this includes the entropy effect in addition
to $\usl$ and $\usv$ as parts of the internal 
energies.~\cite{Surblys2018} 
\section{conclusion}
%
We have given theoretical expressions for the forces exerted on a Wilhelmy plate, which we modeled as a quasi-2D flat and smooth solid plate immersed into a liquid pool of a simple liquid. By a purely mechanical approach, we have derived the expressions for the local forces on the top, the contact line (CL) and the bottom of the plate as well as the total force on the plate. All forces given by the theory showed an excellent agreement with the MD simulation results.

In particular, we have shown that the local force on the CL is written as the difference of the potential energy densities between the SL and SV interfaces away from the CL but not generally as the difference between the SL and SV works of adhesion. On the other hand, we have revealed that the local forces on the top and bottom of the plate can be related to the SV and SL works of adhesion, respectively. As the summation of these local forces, we have obtained the modified form of the Wilhelmy equation, which was consistent with the overall force balance on the system. The modified Wilhelmy equation includes the cofactor taking into account the plate thickness, whose effect can be significant in small systems like the present one. 

Finally, we have shown that with these expressions of the forces all the interfacial tensions $\gsl$ and $\gsv$ as well as $\glv$ can be extracted from a single equilibrium MD simulation without the computationally demanding calculation of the local stress distributions and the thermodynamic integrations.
\begin{acknowledgments}
We thank Konan Imadate  for fruitful discussion. 
T.O.
and Y.Y. are supported by JSPS KAKENHI Grant Nos. JP18K03929
and JP18K03978, Japan, respectively. 
Y.Y. is also supported by JST CREST Grant No. JPMJCR18I1, Japan.
\end{acknowledgments}
\appendix
\section{Relation between the SL interaction parameter and the contact angle
\label{sec:appendix_eta_costheta}
}
\begin{figure}
\includegraphics[width=0.8\linewidth]{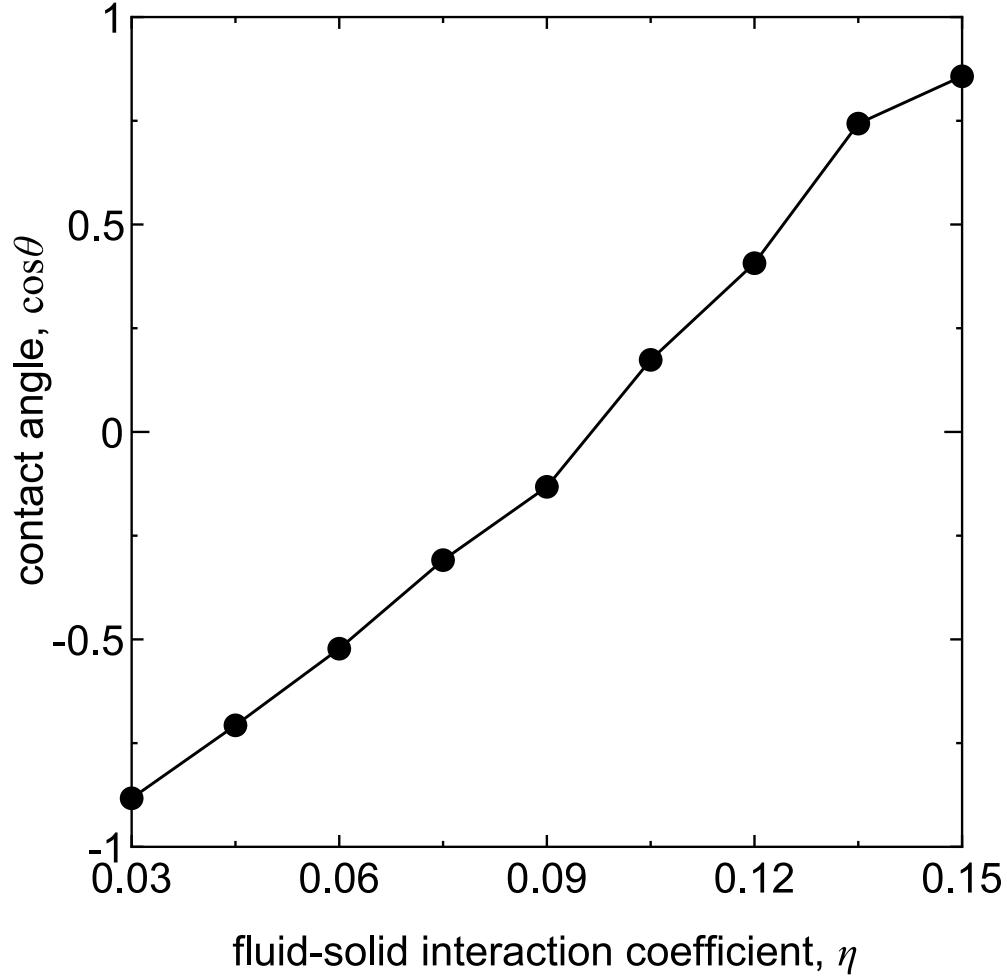}
\caption{Relation between the cosine of the apparent contact angle 
$\cos \theta$ of the meniscus and the SF interaction 
coefficient $\eta$.
\label{Fig:eta-costheta}
}
\end{figure}
In the main text, we summarized the results by $\cos \theta$ as 
the cosine of the apparent contact angle $\theta$ of the meniscus, 
while the SF interaction coefficient $\eta$ was varied as the 
parameter for the MD simulations. As described in the main text, 
we defined $\theta$ by the angle between the SF interface at 
$x=\xsf= 1.15$~nm and the extended cylindrical curved surface of the LV 
interface having a constant curvature determined through the least-squares 
fitting of a circle on the density contour of $\rho=$400~kg/m$^{3}$ 
at the LV interface excluding the region in the adsorption layers 
near the solid surface. 
Figure~\ref{Fig:eta-costheta} shows the relation 
between the SL interaction parameter $\eta$ 
and the apparent contact angle $\theta$.
The contact angle cosine $\cos \theta$ monotonically increased 
with the increase of $\eta$, and a unique relation can be 
obtained between the two for the present range of $\eta$.
\section{Thermodynamic integration (TI) with the dry-surface scheme
\label{sec:appendix_TI}
}
\begin{figure}
\includegraphics[width=0.65\linewidth]{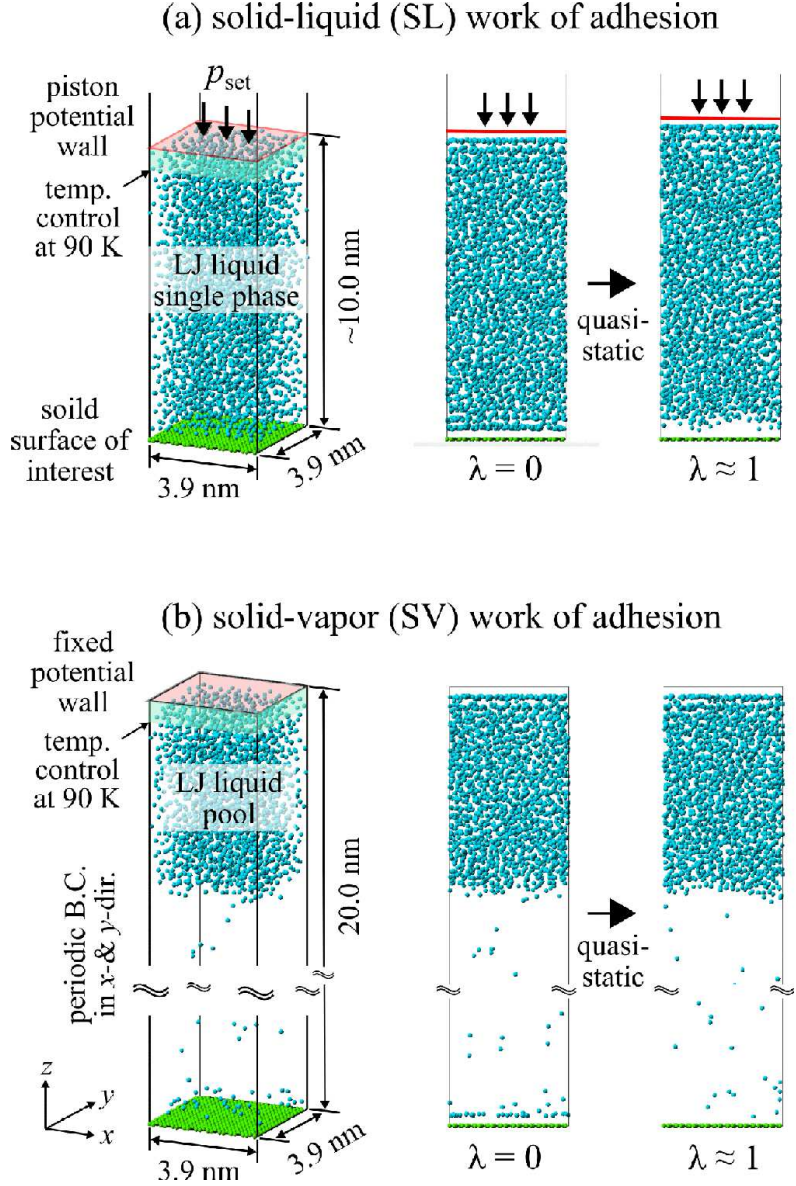}
\caption{Simulation systems for the calculation of the solid-liquid 
and solid-vapor works of adhesion by the thermodynamic integration 
(TI) through the dry-surface (DS) scheme.
\label{Fig:TI-systems}
}
\end{figure}
We calculated the solid-liquid (SL) and solid-vapor (SV) works of 
adhesion $\Wsl$ and $\Wsv$, respectively, by the thermodynamic integration (TI)~\cite{Frenkel2007} 
through the dry-surface (DS) scheme~\cite{Leroy2015} to compare with the 
relative SL and SV interfacial tensions
obtained in the present Wilhelmy MD systems. Details of the DS 
scheme were basically the same as in our previous study.~\cite{Yamaguchi2019} 
In the systems shown in Fig.~\ref{Fig:TI-systems}, 
the liquid or vapor was quasi-statically stripped off from the 
solid surface fixed on the bottom of the coordinate system,
which had the same periodic honeycomb structure as the solid 
plate in the Wilhelmy MD system.
The work of adhesion was calculated as the free energy 
difference after and before the above procedure, where 
the coupling parameter for the TI was embedded 
in the SF interaction parameter in the DS scheme. 
\par
For the calculation of $\Wsl$, a SL interface was formed between 
the liquid and bottom solid as shown in Fig.~\ref{Fig:TI-systems}~(a) 
with wettability parameter $\eta$ corresponding to the Wilhelmy 
MD system.
Periodic boundary condition was employed in the $x$-and $y$-directions tangential to the solid surface. 
In addition, we set a piston at $z=\zpis$ 
above the liquid to attain a constant pressure system.
By allocating sufficient number of fluid particles $N_\mathrm{f}$ 
and by 
setting the pressure $p_\mathrm{set}$ above the vapor pressure, 
a liquid bulk with a constant density was formed between the 
solid wall and piston. We used 3000 fluid particles, and 
the system size was set as shown in Fig.~\ref{Fig:TI-systems}~(a).
We also controlled the temperature of the fluid particles within 
0.8~nm from the top piston regarding the velocity components 
in the $x$- and $y$-directions at $T_\mathrm{c}=90$~K. 
\par
We embedded a coupling parameter $\lambda$ into the SF interaction 
potential given in Eq.~\eqref{eq:LJ} as 
\begin{equation}
  \label{eq:LJcouple}
  \Phi^\mathrm{DS}_\mathrm{sf}(r_{ij},\lambda) = 
  (1-\lambda) \Phi^\mathrm{LJ}_\mathrm{sf}(r_{ij}),
\end{equation}
%
and we obtained multiple equilibrium systems with various $\lambda$ values with $0 \leq \lambda < 1$ to numerically calculate the TI described 
below. Each system was obtained after a preliminary equilibration 
of 10~ns, and the time average of 20~ns was used for the analysis.
\par
The work of adhesion $\Wsl$ is defined by the minimum work needed to strip the liquid from the solid surface per area under constant $NpT$, and it can be calculated by the TI along a reversible path between the initial and final states of the process. In the present DS scheme, this was achieved by at first forming a SL interface, and then by weakening the SF interaction potential through the coupling parameter. We obtained equilibrium SL interfaces with discrete coupling parameter $\lambda$ varied from 0 to 0.999. 
Note that the maximum value of $\lambda$ was set slightly below 1 to keep the SF interaction to be effectively only repulsive. This value is denoted by $1^{-}$ hereafter. 
%
The difference of the SL interfacial Gibbs free energy 
$\Delta G_\mathrm{SL} \equiv 
G_\mathrm{SL}|_{\lambda=1^{-}} - 
G_\mathrm{SL}|_{\lambda=0}$ 
between systems at 
$\lambda=0$ and $\lambda=1^{-}$ under constant $NpT$ 
was related to the difference in the surface 
interfacial energies as 
\begin{eqnarray}
\nonumber
   W_\mathrm{SL} 
   &\equiv&
   \frac{\Delta G_\mathrm{SL}}{A} =
   \gamma_\mathrm{S0} + \gamma_\mathrm{L0} - \gamma_\mathrm{SL}
\\ \label{eq:W_sl_appendix}
  & \approx &
   \gamma_\mathrm{S0} + \gamma_\mathrm{LV} - \gamma_\mathrm{SL},
\end{eqnarray}
where the vacuum phase was denoted by subscript \lq 0' and 
$\gamma_\mathrm{S0}$ and $\gamma_\mathrm{L0}$ were the 
solid-vacuum and liquid-vacuum interfacial energies per unit 
area.
Note that $\gamma_\mathrm{L0}$ was substituted by the liquid-vapor 
interfacial tension $\glv$ in the final approximation considering 
that the vapor density was negligibly small.
Using the $NpT$ canonical ensemble, the difference of 
the SL interfacial Gibbs free energy 
$\Delta G_\mathrm{SL}$ in Eq.~\eqref{eq:W_sl_appendix} 
was calculated through the following TI:
\begin{eqnarray}
\Delta G
 &=&
\int_0^{1^{-}} \frac{d G(\lambda)}{d \lambda} d \lambda
	\nonumber
=
\int_0^{1^{-}} \angb{
  \frac{\partial H}{\partial \lambda} 
} d \lambda
\nonumber
\\
&=&
-\int_0^{1^{-}} \angb{
 \sum_{i\in\mathrm{fluid}}^{N_\mathrm{f}} \sum_{j\in\mathrm{wall}}^{N_\mathrm{w}}
  \Phi_\mathrm{fw}
} d \lambda,
\end{eqnarray}
\begin{equation}
	\Delta G_\mathrm{SL}  = 
	\Delta G - 
	A p_\mathrm{set} \left( 
	\angb{ z_\mathrm{p} |_{\lambda =1^{-}} } 
	- 
	\angb{ z_\mathrm{p} |_{\lambda =0} }
	\right)
	\label{eq:DeltaGSL=DeltaG-work_piston}
\end{equation}
where $H$ was the  Hamiltonian, \ie
the internal energy of the system
and $N_\mathrm{w}$ was the numbers of wall molecules. 
The ensemble average was substituted by the time average 
in the simulation, and was denoted by the angle brackets.
Note that to obtain $\Delta G_\mathrm{SL}$,  
the work exerted on the piston 
$A p_\mathrm{set} \left( 
\angb{ z_\mathrm{p} |_{\lambda =1^{-}} } 
- 
\angb{ z_\mathrm{p} |_{\lambda =0} }
\right)$
was subtracted from the change of 
the Gibbs free energy of the system $\Delta G$ 
including the piston in Eq.~\eqref{eq:DeltaGSL=DeltaG-work_piston}.
\par
For the calculation of the SV work of adhesion $\Wsv$, we investigated the interfacial energy between saturated vapor and corresponding solid surface set on the bottom of the simulation cell by placing an additional particle bath on the top as shown in Fig.~\ref{Fig:TI-systems}~(b). The setup regarding the periodic boundary conditions employed in $x$-and $y$-directions, temperature control and placement conditions for the solid surface were the same as the SL system, whereas the particle bath was kept in place by a potential field 
at a fixed height sufficiently far from the solid surface. 
This potential field mimicked a completely wettable surface with an equilibrium contact angle of zero with the present potential parameters, \ie a liquid film was formed on the particle bath. With this setting, a solid-vapor interface with the same density distribution as that in the 
Wilhelmy MD system was achieved. We formed multiple equilibrium systems with various values of the coupling parameter $\lambda$ with the same recipe as the SL systems.
\par
Similar to the calculation of $\Wsl$, the SV interface at $\lambda=0$
was divided into S0 and V0 interfaces at $\lambda = 1^{-}$ 
as shown in Fig.~\ref{Fig:TI-systems}~(b), while the calculation
systems for $\Wsv$ were under constant $NVT$. Thus, the solid-vapor 
work of adhesion $W_\mathrm{SV}$ was given by the 
difference of the Helmholtz free energy $\Delta F$ 
per unit area, and was related to the difference in the 
surface interfacial energy as 
\begin{eqnarray}
\nonumber
   W_\mathrm{SV} 
   &\equiv&
   \frac{\Delta{F}}{A} =
   \gamma_\mathrm{S0} + \gamma_\mathrm{V0} - \gamma_\mathrm{SV}
\\ \label{eq:W_sv_appendix}
  & \approx &
   \gamma_\mathrm{S0} - \gamma_\mathrm{SV},
\end{eqnarray}
where $\gamma_\mathrm{V0}$ was set zero in the final 
approximation. 
Using the $NVT$ canonical ensemble, $\Delta F$ in 
Eq.~\eqref{eq:W_sv_appendix} was calculated through the TI as:
\begin{eqnarray}
   \nonumber
   \Delta{F} &=&
   \int_0^{1^{-}} \frac{\partial F(\lambda)}{\partial \lambda} d \lambda
   =\int_0^{1^{-}} 
   \angb{ \frac{\partial H}{\partial \lambda} } d \lambda
   \\
   \label{eq:thermo_sv}
   &=&
   -\int_0^{1^{-}} \angb{
   \sum_{i}^{N_\mathrm{f}} \sum_{j}^{N_\mathrm{w} } 
   \Phi^\mathrm{LJ}_\mathrm{fw}(r_{ij})  
   } d \lambda.
\end{eqnarray}
\par
\begin{figure}
\includegraphics[width=0.8\linewidth]{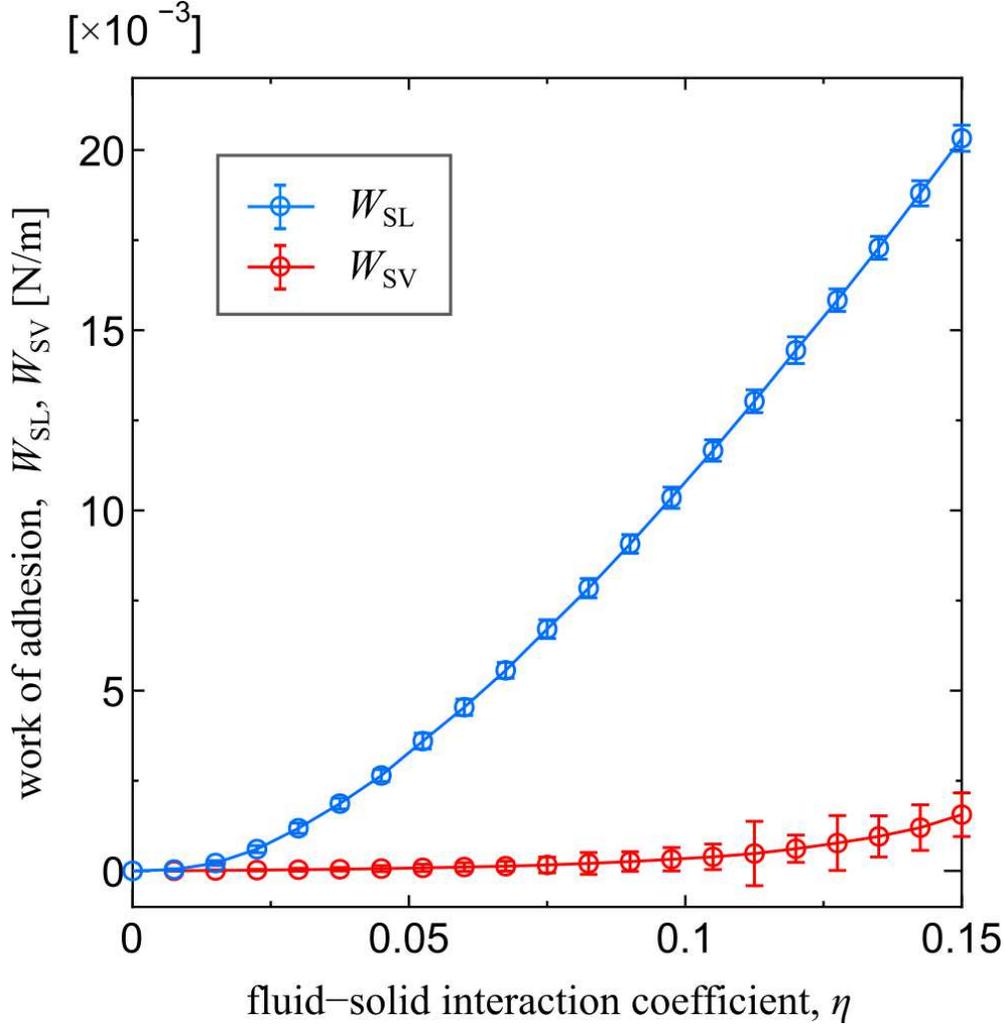}
\caption{Works of adhesion $\Wsl$ and $\Wsv$ calculated by the 
TI as a function of the solid-fluid interaction coefficient $\eta$.
\label{Fig:TI-results}
}
\end{figure}
Figure~\ref{Fig:TI-results} shows the SL and SV works 
of adhesion $\Wsl$ and $\Wsv$ calculated by the TI as a 
function of the solid-fluid interaction coefficient $\eta$.
These values were used for the results shown in  Fig.~\ref{Fig:fig08-bot-top-Wil-TI} through $\eta$-$\cos \theta$ 
relation in Fig.~\ref{Fig:eta-costheta}.
\newline
\par \noindent
\textbf{DATA AVAILABILITY}
\newline
\par
The data that support the findings of this study are available from the corresponding author
upon reasonable request.
%

\begin{thebibliography}{53}%
\makeatletter
\providecommand \@ifxundefined [1]{%
 \@ifx{#1\undefined}
}%
\providecommand \@ifnum [1]{%
 \ifnum #1\expandafter \@firstoftwo
 \else \expandafter \@secondoftwo
 \fi
}%
\providecommand \@ifx [1]{%
 \ifx #1\expandafter \@firstoftwo
 \else \expandafter \@secondoftwo
 \fi
}%
\providecommand \natexlab [1]{#1}%
\providecommand \enquote  [1]{``#1''}%
\providecommand \bibnamefont  [1]{#1}%
\providecommand \bibfnamefont [1]{#1}%
\providecommand \citenamefont [1]{#1}%
\providecommand \href@noop [0]{\@secondoftwo}%
\providecommand \href [0]{\begingroup \@sanitize@url \@href}%
\providecommand \@href[1]{\@@startlink{#1}\@@href}%
\providecommand \@@href[1]{\endgroup#1\@@endlink}%
\providecommand \@sanitize@url [0]{\catcode `\\12\catcode `\$12\catcode
  `\&12\catcode `\#12\catcode `\^12\catcode `\_12\catcode `\%12\relax}%
\providecommand \@@startlink[1]{}%
\providecommand \@@endlink[0]{}%
\providecommand \url  [0]{\begingroup\@sanitize@url \@url }%
\providecommand \@url [1]{\endgroup\@href {#1}{\urlprefix }}%
\providecommand \urlprefix  [0]{URL }%
\providecommand \Eprint [0]{\href }%
\providecommand \doibase [0]{http://dx.doi.org/}%
\providecommand \selectlanguage [0]{\@gobble}%
\providecommand \bibinfo  [0]{\@secondoftwo}%
\providecommand \bibfield  [0]{\@secondoftwo}%
\providecommand \translation [1]{[#1]}%
\providecommand \BibitemOpen [0]{}%
\providecommand \bibitemStop [0]{}%
\providecommand \bibitemNoStop [0]{.\EOS\space}%
\providecommand \EOS [0]{\spacefactor3000\relax}%
\providecommand \BibitemShut  [1]{\csname bibitem#1\endcsname}%
\let\auto@bib@innerbib\@empty
\bibitem [{\citenamefont {de~Gennes}(1985)}]{deGenne1985}%
  \BibitemOpen
  \bibfield  {author} {\bibinfo {author} {\bibfnamefont {P.-G.}\ \bibnamefont
  {de~Gennes}},\ }\bibfield  {title} {\enquote {\bibinfo {title} {{``Wetting''
  Statics and dynamics}},}\ }\href@noop {} {\bibfield  {journal} {\bibinfo
  {journal} {Rev. Mod. Phys.}\ }\textbf {\bibinfo {volume} {57}},\ \bibinfo
  {pages} {827--863} (\bibinfo {year} {1985})}\BibitemShut {NoStop}%
\bibitem [{\citenamefont {Ono}\ and\ \citenamefont {Kondo}(1960)}]{Ono1960}%
  \BibitemOpen
  \bibfield  {author} {\bibinfo {author} {\bibfnamefont {S.}~\bibnamefont
  {Ono}}\ and\ \bibinfo {author} {\bibfnamefont {S.}~\bibnamefont {Kondo}},\
  }\href@noop {} {\emph {\bibinfo {title} {Molecular Theory of Surface Tension
  in Liquids}}},\ Encyclopedia of Physics / Handbuch der Physik\ (\bibinfo
  {publisher} {Springer},\ \bibinfo {year} {1960})\ pp.\ \bibinfo {pages}
  {134--280}\BibitemShut {NoStop}%
\bibitem [{\citenamefont {Rowlinson}\ and\ \citenamefont
  {Widom}(1982)}]{Rowlinson1982}%
  \BibitemOpen
  \bibfield  {author} {\bibinfo {author} {\bibfnamefont {J.~S.}\ \bibnamefont
  {Rowlinson}}\ and\ \bibinfo {author} {\bibfnamefont {B.}~\bibnamefont
  {Widom}},\ }\href@noop {} {\emph {\bibinfo {title} {Molecular Theory of
  Capillarity}}}\ (\bibinfo  {publisher} {Dover},\ \bibinfo {year}
  {1982})\BibitemShut {NoStop}%
\bibitem [{\citenamefont {Schimmele}, \citenamefont {Napl{\'{o}}rkowski},\ and\
  \citenamefont {Dietrich}(2007)}]{Schimmele2007}%
  \BibitemOpen
  \bibfield  {author} {\bibinfo {author} {\bibfnamefont {L.}~\bibnamefont
  {Schimmele}}, \bibinfo {author} {\bibfnamefont {M.}~\bibnamefont
  {Napl{\'{o}}rkowski}}, \ and\ \bibinfo {author} {\bibfnamefont
  {S.}~\bibnamefont {Dietrich}},\ }\bibfield  {title} {\enquote {\bibinfo
  {title} {{Conceptual aspects of line tensions}},}\ }\href {\doibase
  10.1063/1.2799990} {\bibfield  {journal} {\bibinfo  {journal} {J. Chem.
  Phys.}\ }\textbf {\bibinfo {volume} {127}} (\bibinfo {year} {2007}),\
  10.1063/1.2799990},\ \Eprint {http://arxiv.org/abs/0703821} {0703821
  [cond-mat]} \BibitemShut {NoStop}%
\bibitem [{\citenamefont {Drelich}\ \emph {et~al.}(2019)\citenamefont
  {Drelich}, \citenamefont {Boinovich}, \citenamefont {Chibowski},
  \citenamefont {Volpe}, \citenamefont {Ho{\l}ysz}, \citenamefont {Marmur},\
  and\ \citenamefont {Siboni}}]{Drelich2019}%
  \BibitemOpen
  \bibfield  {author} {\bibinfo {author} {\bibfnamefont {J.~W.}\ \bibnamefont
  {Drelich}}, \bibinfo {author} {\bibfnamefont {L.}~\bibnamefont {Boinovich}},
  \bibinfo {author} {\bibfnamefont {E.}~\bibnamefont {Chibowski}}, \bibinfo
  {author} {\bibfnamefont {C.~D.}\ \bibnamefont {Volpe}}, \bibinfo {author}
  {\bibfnamefont {L.}~\bibnamefont {Ho{\l}ysz}}, \bibinfo {author}
  {\bibfnamefont {A.}~\bibnamefont {Marmur}}, \ and\ \bibinfo {author}
  {\bibfnamefont {S.}~\bibnamefont {Siboni}},\ }\bibfield  {title} {\enquote
  {\bibinfo {title} {Contact angles: History of over 200 years of open
  questions},}\ }\href {\doibase 10.1680/jsuin.19.00007} {\bibfield  {journal}
  {\bibinfo  {journal} {Surf. Innov.}\ ,\ \bibinfo {pages} {1--25}} (\bibinfo
  {year} {2019})}\BibitemShut {NoStop}%
\bibitem [{\citenamefont {Young}(1805)}]{Young1805}%
  \BibitemOpen
  \bibfield  {author} {\bibinfo {author} {\bibfnamefont {T.}~\bibnamefont
  {Young}},\ }\bibfield  {title} {\enquote {\bibinfo {title} {An essay on the
  cohesion of fluids},}\ }\href {\doibase 10.1098/rstl.1805.0005} {\bibfield
  {journal} {\bibinfo  {journal} {Phil. Trans. R. Soc. Lond.}\ }\textbf
  {\bibinfo {volume} {95}},\ \bibinfo {pages} {65} (\bibinfo {year}
  {1805})}\BibitemShut {NoStop}%
\bibitem [{\citenamefont {Gao}\ and\ \citenamefont {McCarthy}(2009)}]{Gao2009}%
  \BibitemOpen
  \bibfield  {author} {\bibinfo {author} {\bibfnamefont {L.}~\bibnamefont
  {Gao}}\ and\ \bibinfo {author} {\bibfnamefont {T.~J.}\ \bibnamefont
  {McCarthy}},\ }\bibfield  {title} {\enquote {\bibinfo {title} {Wetting
  101$^{\circ}$},}\ }\href@noop {} {\bibfield  {journal} {\bibinfo  {journal}
  {Langmuir}\ }\textbf {\bibinfo {volume} {25}},\ \bibinfo {pages}
  {14105--14115} (\bibinfo {year} {2009})}\BibitemShut {NoStop}%
\bibitem [{\citenamefont {Tanaka}, \citenamefont {Morigami},\ and\
  \citenamefont {Atoda}(1993)}]{Tanaka1993}%
  \BibitemOpen
  \bibfield  {author} {\bibinfo {author} {\bibfnamefont {T.}~\bibnamefont
  {Tanaka}}, \bibinfo {author} {\bibfnamefont {M.}~\bibnamefont {Morigami}}, \
  and\ \bibinfo {author} {\bibfnamefont {N.}~\bibnamefont {Atoda}},\ }\bibfield
   {title} {\enquote {\bibinfo {title} {Mechanism of resist pattern collapse
  during development process},}\ }\href {\doibase 10.1143/jjap.32.6059}
  {\bibfield  {journal} {\bibinfo  {journal} {Jap. J. Appl. Phys.}\ }\textbf
  {\bibinfo {volume} {32}},\ \bibinfo {pages} {6059--6064} (\bibinfo {year}
  {1993})}\BibitemShut {NoStop}%
\bibitem [{\citenamefont {Kirkwood}\ and\ \citenamefont
  {Buff}(1949)}]{Kirkwood1949}%
  \BibitemOpen
  \bibfield  {author} {\bibinfo {author} {\bibfnamefont {J.~G.}\ \bibnamefont
  {Kirkwood}}\ and\ \bibinfo {author} {\bibfnamefont {F.~P.}\ \bibnamefont
  {Buff}},\ }\bibfield  {title} {\enquote {\bibinfo {title} {The statistical
  mechanical theory of surface tension},}\ }\href {\doibase 10.1063/1.1747248}
  {\bibfield  {journal} {\bibinfo  {journal} {J. Chem. Phys.}\ }\textbf
  {\bibinfo {volume} {17}},\ \bibinfo {pages} {338--343} (\bibinfo {year}
  {1949})}\BibitemShut {NoStop}%
\bibitem [{\citenamefont {Nijmeijer}\ and\ \citenamefont {van
  Leeuwen}(1990)}]{Nijmeijer1990_theor}%
  \BibitemOpen
  \bibfield  {author} {\bibinfo {author} {\bibfnamefont {M.~J.~P.}\
  \bibnamefont {Nijmeijer}}\ and\ \bibinfo {author} {\bibfnamefont {J.~M.~J.}\
  \bibnamefont {van Leeuwen}},\ }\bibfield  {title} {\enquote {\bibinfo {title}
  {Microscopic expressions for the surface and line tension},}\ }\href@noop {}
  {\bibfield  {journal} {\bibinfo  {journal} {J. Phys. A: Math. Gen.}\ }\textbf
  {\bibinfo {volume} {23}},\ \bibinfo {pages} {4211--4235} (\bibinfo {year}
  {1990})}\BibitemShut {NoStop}%
\bibitem [{\citenamefont {Nijmeijer}\ \emph {et~al.}(1990)\citenamefont
  {Nijmeijer}, \citenamefont {Bruin}, \citenamefont {Bakker},\ and\
  \citenamefont {van Leeuwen}}]{Nijmeijer1990_simul}%
  \BibitemOpen
  \bibfield  {author} {\bibinfo {author} {\bibfnamefont {M.~J.~P.}\
  \bibnamefont {Nijmeijer}}, \bibinfo {author} {\bibfnamefont {C.}~\bibnamefont
  {Bruin}}, \bibinfo {author} {\bibfnamefont {A.~F.}\ \bibnamefont {Bakker}}, \
  and\ \bibinfo {author} {\bibfnamefont {J.~M.~J.}\ \bibnamefont {van
  Leeuwen}},\ }\bibfield  {title} {\enquote {\bibinfo {title} {Wetting and
  drying of an inert wall by a fluid in a molecular-dynamics simulation},}\
  }\href@noop {} {\bibfield  {journal} {\bibinfo  {journal} {Phys. Rev. A}\
  }\textbf {\bibinfo {volume} {42}},\ \bibinfo {pages} {6052--6059} (\bibinfo
  {year} {1990})}\BibitemShut {NoStop}%
\bibitem [{\citenamefont {Tang}\ and\ \citenamefont {Harris}(1995)}]{Tang1995}%
  \BibitemOpen
  \bibfield  {author} {\bibinfo {author} {\bibfnamefont {J.~Z.}\ \bibnamefont
  {Tang}}\ and\ \bibinfo {author} {\bibfnamefont {J.~G.}\ \bibnamefont
  {Harris}},\ }\bibfield  {title} {\enquote {\bibinfo {title} {Fluid wetting on
  molecularly rough surfaces},}\ }\href@noop {} {\bibfield  {journal} {\bibinfo
   {journal} {J. Chem. Phys.}\ }\textbf {\bibinfo {volume} {103}},\ \bibinfo
  {pages} {8201--8208} (\bibinfo {year} {1995})}\BibitemShut {NoStop}%
\bibitem [{\citenamefont {Gloor}\ \emph {et~al.}(2005)\citenamefont {Gloor},
  \citenamefont {Jackson}, \citenamefont {Blas},\ and\ \citenamefont
  {De~Miguel}}]{Gloor2005}%
  \BibitemOpen
  \bibfield  {author} {\bibinfo {author} {\bibfnamefont {G.~J.}\ \bibnamefont
  {Gloor}}, \bibinfo {author} {\bibfnamefont {G.}~\bibnamefont {Jackson}},
  \bibinfo {author} {\bibfnamefont {F.~J.}\ \bibnamefont {Blas}}, \ and\
  \bibinfo {author} {\bibfnamefont {E.}~\bibnamefont {De~Miguel}},\ }\bibfield
  {title} {\enquote {\bibinfo {title} {Test-area simulation method for the
  direct determination of the interfacial tension of systems with continuous or
  discontinuous potentials},}\ }\href@noop {} {\bibfield  {journal} {\bibinfo
  {journal} {J. Chem. Phys.}\ }\textbf {\bibinfo {volume} {123}},\ \bibinfo
  {pages} {134703} (\bibinfo {year} {2005})}\BibitemShut {NoStop}%
\bibitem [{\citenamefont {Ingebrigtsen}\ and\ \citenamefont
  {Toxvaerd}(2007)}]{Ingebrigtsen2007}%
  \BibitemOpen
  \bibfield  {author} {\bibinfo {author} {\bibfnamefont {T.}~\bibnamefont
  {Ingebrigtsen}}\ and\ \bibinfo {author} {\bibfnamefont {S.}~\bibnamefont
  {Toxvaerd}},\ }\bibfield  {title} {\enquote {\bibinfo {title} {Contact angles
  of {Lennard-Jones} liquids and droplets on planar surfaces},}\ }\href@noop {}
  {\bibfield  {journal} {\bibinfo  {journal} {J. Phys. Chem. C}\ }\textbf
  {\bibinfo {volume} {111}},\ \bibinfo {pages} {8518--8523} (\bibinfo {year}
  {2007})}\BibitemShut {NoStop}%
\bibitem [{\citenamefont {Das}\ and\ \citenamefont {Binder}(2010)}]{Das2010}%
  \BibitemOpen
  \bibfield  {author} {\bibinfo {author} {\bibfnamefont {S.~K.}\ \bibnamefont
  {Das}}\ and\ \bibinfo {author} {\bibfnamefont {K.}~\bibnamefont {Binder}},\
  }\bibfield  {title} {\enquote {\bibinfo {title} {{Does Young's equation hold
  on the nanoscale? A Monte Carlo test for the binary Lennard-Jones fluid}},}\
  }\href@noop {} {\bibfield  {journal} {\bibinfo  {journal} {Europhy. Lett.}\
  }\textbf {\bibinfo {volume} {92}},\ \bibinfo {pages} {26006} (\bibinfo {year}
  {2010})}\BibitemShut {NoStop}%
\bibitem [{\citenamefont {Weijs}\ \emph {et~al.}(2011)\citenamefont {Weijs},
  \citenamefont {Marchand}, \citenamefont {Andreotti}, \citenamefont {Lohse},\
  and\ \citenamefont {Snoeijer}}]{Weijs2011}%
  \BibitemOpen
  \bibfield  {author} {\bibinfo {author} {\bibfnamefont {J.~H.}\ \bibnamefont
  {Weijs}}, \bibinfo {author} {\bibfnamefont {A.}~\bibnamefont {Marchand}},
  \bibinfo {author} {\bibfnamefont {B.}~\bibnamefont {Andreotti}}, \bibinfo
  {author} {\bibfnamefont {D.}~\bibnamefont {Lohse}}, \ and\ \bibinfo {author}
  {\bibfnamefont {J.~H.}\ \bibnamefont {Snoeijer}},\ }\bibfield  {title}
  {\enquote {\bibinfo {title} {Origin of line tension for a {Lennard-Jones}
  nanodroplet},}\ }\href@noop {} {\bibfield  {journal} {\bibinfo  {journal}
  {Phys. Fluids}\ }\textbf {\bibinfo {volume} {23}},\ \bibinfo {pages} {022001}
  (\bibinfo {year} {2011})}\BibitemShut {NoStop}%
\bibitem [{\citenamefont {Seveno}, \citenamefont {Blake},\ and\ \citenamefont
  {de~Coninck}(2013)}]{Seveno2013}%
  \BibitemOpen
  \bibfield  {author} {\bibinfo {author} {\bibfnamefont {D.}~\bibnamefont
  {Seveno}}, \bibinfo {author} {\bibfnamefont {T.~D.}\ \bibnamefont {Blake}}, \
  and\ \bibinfo {author} {\bibfnamefont {J.}~\bibnamefont {de~Coninck}},\
  }\bibfield  {title} {\enquote {\bibinfo {title} {Young's equation at the
  nanoscale},}\ }\href {\doibase 10.1103/PhysRevLett.111.096101} {\bibfield
  {journal} {\bibinfo  {journal} {Phys. Rev. Lett.}\ }\textbf {\bibinfo
  {volume} {111}},\ \bibinfo {pages} {096101} (\bibinfo {year}
  {2013})}\BibitemShut {NoStop}%
\bibitem [{\citenamefont {Surblys}\ \emph {et~al.}(2014)\citenamefont
  {Surblys}, \citenamefont {Yamaguchi}, \citenamefont {Kuroda}, \citenamefont
  {Kagawa}, \citenamefont {Nakajima},\ and\ \citenamefont
  {Fujimura}}]{Surblys2014}%
  \BibitemOpen
  \bibfield  {author} {\bibinfo {author} {\bibfnamefont {D.}~\bibnamefont
  {Surblys}}, \bibinfo {author} {\bibfnamefont {Y.}~\bibnamefont {Yamaguchi}},
  \bibinfo {author} {\bibfnamefont {K.}~\bibnamefont {Kuroda}}, \bibinfo
  {author} {\bibfnamefont {M.}~\bibnamefont {Kagawa}}, \bibinfo {author}
  {\bibfnamefont {T.}~\bibnamefont {Nakajima}}, \ and\ \bibinfo {author}
  {\bibfnamefont {H.}~\bibnamefont {Fujimura}},\ }\bibfield  {title} {\enquote
  {\bibinfo {title} {Molecular dynamics analysis on wetting and interfacial
  properties of water-alcohol mixture droplets on a solid surface},}\ }\href
  {\doibase http://dx.doi.org/10.1063/1.4861039} {\bibfield  {journal}
  {\bibinfo  {journal} {J. Chem. Phys.}\ }\textbf {\bibinfo {volume} {140}},\
  \bibinfo {eid} {034505} (\bibinfo {year} {2014})}\BibitemShut {NoStop}%
\bibitem [{\citenamefont {Nishida}\ \emph {et~al.}(2014)\citenamefont
  {Nishida}, \citenamefont {Surblys}, \citenamefont {Yamaguchi}, \citenamefont
  {Kuroda}, \citenamefont {Kagawa}, \citenamefont {Nakajima},\ and\
  \citenamefont {Fujimura}}]{Nishida2014}%
  \BibitemOpen
  \bibfield  {author} {\bibinfo {author} {\bibfnamefont {S.}~\bibnamefont
  {Nishida}}, \bibinfo {author} {\bibfnamefont {D.}~\bibnamefont {Surblys}},
  \bibinfo {author} {\bibfnamefont {Y.}~\bibnamefont {Yamaguchi}}, \bibinfo
  {author} {\bibfnamefont {K.}~\bibnamefont {Kuroda}}, \bibinfo {author}
  {\bibfnamefont {M.}~\bibnamefont {Kagawa}}, \bibinfo {author} {\bibfnamefont
  {T.}~\bibnamefont {Nakajima}}, \ and\ \bibinfo {author} {\bibfnamefont
  {H.}~\bibnamefont {Fujimura}},\ }\bibfield  {title} {\enquote {\bibinfo
  {title} {Molecular dynamics analysis of multiphase interfaces based on
  {\textit{in situ}} extraction of the pressure distribution of a liquid
  droplet on a solid surface},}\ }\href {\doibase
  http://dx.doi.org/10.1063/1.4865254} {\bibfield  {journal} {\bibinfo
  {journal} {J. Chem. Phys.}\ }\textbf {\bibinfo {volume} {140}},\ \bibinfo
  {pages} {074707} (\bibinfo {year} {2014})}\BibitemShut {NoStop}%
\bibitem [{\citenamefont {Lau}\ \emph {et~al.}(2015)\citenamefont {Lau},
  \citenamefont {Ford}, \citenamefont {Hunt}, \citenamefont {M{\"u}ller},\ and\
  \citenamefont {Jackson}}]{Lau2015}%
  \BibitemOpen
  \bibfield  {author} {\bibinfo {author} {\bibfnamefont {G.~V.}\ \bibnamefont
  {Lau}}, \bibinfo {author} {\bibfnamefont {I.~J.}\ \bibnamefont {Ford}},
  \bibinfo {author} {\bibfnamefont {P.~A.}\ \bibnamefont {Hunt}}, \bibinfo
  {author} {\bibfnamefont {E.~A.}\ \bibnamefont {M{\"u}ller}}, \ and\ \bibinfo
  {author} {\bibfnamefont {G.}~\bibnamefont {Jackson}},\ }\bibfield  {title}
  {\enquote {\bibinfo {title} {Surface thermodynamics of planar, cylindrical,
  and spherical vapour-liquid interfaces of water},}\ }\href@noop {} {\bibfield
   {journal} {\bibinfo  {journal} {J. Chem. Phys.}\ }\textbf {\bibinfo {volume}
  {142}},\ \bibinfo {pages} {114701} (\bibinfo {year} {2015})}\BibitemShut
  {NoStop}%
\bibitem [{\citenamefont {Yamaguchi}\ \emph {et~al.}(2019)\citenamefont
  {Yamaguchi}, \citenamefont {Kusudo}, \citenamefont {Surblys}, \citenamefont
  {Omori},\ and\ \citenamefont {Kikugawa}}]{Yamaguchi2019}%
  \BibitemOpen
  \bibfield  {author} {\bibinfo {author} {\bibfnamefont {Y.}~\bibnamefont
  {Yamaguchi}}, \bibinfo {author} {\bibfnamefont {H.}~\bibnamefont {Kusudo}},
  \bibinfo {author} {\bibfnamefont {D.}~\bibnamefont {Surblys}}, \bibinfo
  {author} {\bibfnamefont {T.}~\bibnamefont {Omori}}, \ and\ \bibinfo {author}
  {\bibfnamefont {G.}~\bibnamefont {Kikugawa}},\ }\bibfield  {title} {\enquote
  {\bibinfo {title} {{Interpretation of Young's equation for a liquid droplet
  on a flat and smooth solid surface: Mechanical and thermodynamic routes with
  a simple Lennard-Jones liquid}},}\ }\href {\doibase 10.1063/1.5053881}
  {\bibfield  {journal} {\bibinfo  {journal} {J. Chem. Phys.}\ }\textbf
  {\bibinfo {volume} {150}},\ \bibinfo {pages} {044701} (\bibinfo {year}
  {2019})}\BibitemShut {NoStop}%
\bibitem [{\citenamefont {Kusudo}, \citenamefont {Omori},\ and\ \citenamefont
  {Yamaguchi}(2019)}]{Kusudo2019}%
  \BibitemOpen
  \bibfield  {author} {\bibinfo {author} {\bibfnamefont {H.}~\bibnamefont
  {Kusudo}}, \bibinfo {author} {\bibfnamefont {T.}~\bibnamefont {Omori}}, \
  and\ \bibinfo {author} {\bibfnamefont {Y.}~\bibnamefont {Yamaguchi}},\
  }\bibfield  {title} {\enquote {\bibinfo {title} {{Extraction of the
  equilibrium pinning force on a contact line exerted from a wettability
  boundary of a solid surface through the connection between mechanical and
  thermodynamic routes}},}\ }\href {\doibase 10.1063/1.5124014} {\bibfield
  {journal} {\bibinfo  {journal} {J. Chem. Phys.}\ }\textbf {\bibinfo {volume}
  {151}},\ \bibinfo {pages} {154501} (\bibinfo {year} {2019})}\BibitemShut
  {NoStop}%
\bibitem [{\citenamefont {Bey}, \citenamefont {Coasne},\ and\ \citenamefont
  {Picard}(2020)}]{Bey2020}%
  \BibitemOpen
  \bibfield  {author} {\bibinfo {author} {\bibfnamefont {R.}~\bibnamefont
  {Bey}}, \bibinfo {author} {\bibfnamefont {B.}~\bibnamefont {Coasne}}, \ and\
  \bibinfo {author} {\bibfnamefont {C.}~\bibnamefont {Picard}},\ }\bibfield
  {title} {\enquote {\bibinfo {title} {{Probing the concept of line tension
  down to the nanoscale}},}\ }\href {\doibase 10.1063/1.5143201} {\bibfield
  {journal} {\bibinfo  {journal} {J Chem. Phys.}\ }\textbf {\bibinfo {volume}
  {152}},\ \bibinfo {pages} {094707} (\bibinfo {year} {2020})}\BibitemShut
  {NoStop}%
\bibitem [{\citenamefont {Grzelak}\ and\ \citenamefont
  {Errington}(2008)}]{Grzelak2008}%
  \BibitemOpen
  \bibfield  {author} {\bibinfo {author} {\bibfnamefont {E.~M.}\ \bibnamefont
  {Grzelak}}\ and\ \bibinfo {author} {\bibfnamefont {J.~R.}\ \bibnamefont
  {Errington}},\ }\bibfield  {title} {\enquote {\bibinfo {title} {Computation
  of interfacial properties via grand canonical transition matrix monte carlo
  simulation},}\ }\href@noop {} {\bibfield  {journal} {\bibinfo  {journal} {J.
  Chem. Phys.}\ }\textbf {\bibinfo {volume} {128}},\ \bibinfo {pages} {014710}
  (\bibinfo {year} {2008})}\BibitemShut {NoStop}%
\bibitem [{\citenamefont {Leroy}, \citenamefont {Dos~Santos},\ and\
  \citenamefont {M{\"u}ller-Plathe}(2009)}]{Leroy2009}%
  \BibitemOpen
  \bibfield  {author} {\bibinfo {author} {\bibfnamefont {F.}~\bibnamefont
  {Leroy}}, \bibinfo {author} {\bibfnamefont {D.~J. V.~A.}\ \bibnamefont
  {Dos~Santos}}, \ and\ \bibinfo {author} {\bibfnamefont {F.}~\bibnamefont
  {M{\"u}ller-Plathe}},\ }\bibfield  {title} {\enquote {\bibinfo {title}
  {Interfacial excess free energies of solid-liquid interfaces by molecular
  dynamics simulation and thermodynamic integration},}\ }\href@noop {}
  {\bibfield  {journal} {\bibinfo  {journal} {Macromol. Rapid Commun.}\
  }\textbf {\bibinfo {volume} {30}},\ \bibinfo {pages} {864--870} (\bibinfo
  {year} {2009})}\BibitemShut {NoStop}%
\bibitem [{\citenamefont {Leroy}\ and\ \citenamefont
  {M{\"u}ller-Plathe}(2010)}]{Leroy2010}%
  \BibitemOpen
  \bibfield  {author} {\bibinfo {author} {\bibfnamefont {F.}~\bibnamefont
  {Leroy}}\ and\ \bibinfo {author} {\bibfnamefont {F.}~\bibnamefont
  {M{\"u}ller-Plathe}},\ }\bibfield  {title} {\enquote {\bibinfo {title}
  {Solid-liquid surface free energy of {Lennard-Jones} liquid on smooth and
  rough surfaces computed by molecular dynamics using the phantom-wall
  method},}\ }\href@noop {} {\bibfield  {journal} {\bibinfo  {journal} {J.
  Chem. Phys.}\ }\textbf {\bibinfo {volume} {133}},\ \bibinfo {pages} {044110}
  (\bibinfo {year} {2010})}\BibitemShut {NoStop}%
\bibitem [{\citenamefont {Kumar}\ and\ \citenamefont
  {Errington}(2014)}]{Kumar2014}%
  \BibitemOpen
  \bibfield  {author} {\bibinfo {author} {\bibfnamefont {B.}~\bibnamefont
  {Kumar}}\ and\ \bibinfo {author} {\bibfnamefont {J.~R.}\ \bibnamefont
  {Errington}},\ }\bibfield  {title} {\enquote {\bibinfo {title} {The use of
  monte carlo simulation to obtain the wetting properties of water},}\
  }\href@noop {} {\bibfield  {journal} {\bibinfo  {journal} {Physics Procedia}\
  }\textbf {\bibinfo {volume} {53}},\ \bibinfo {pages} {44--49} (\bibinfo
  {year} {2014})}\BibitemShut {NoStop}%
\bibitem [{\citenamefont {Leroy}\ and\ \citenamefont
  {M{\"u}ller-Plathe}(2015)}]{Leroy2015}%
  \BibitemOpen
  \bibfield  {author} {\bibinfo {author} {\bibfnamefont {F.}~\bibnamefont
  {Leroy}}\ and\ \bibinfo {author} {\bibfnamefont {F.}~\bibnamefont
  {M{\"u}ller-Plathe}},\ }\bibfield  {title} {\enquote {\bibinfo {title}
  {Dry-surface simulation method for the determination of the work of adhesion
  of solid–liquid interfaces},}\ }\href {\doibase
  10.1021/acs.langmuir.5b01394} {\bibfield  {journal} {\bibinfo  {journal}
  {Langmuir}\ }\textbf {\bibinfo {volume} {31}},\ \bibinfo {pages}
  {8335--–8345} (\bibinfo {year} {2015})}\BibitemShut {NoStop}%
\bibitem [{\citenamefont {Ardham}\ \emph {et~al.}(2015)\citenamefont {Ardham},
  \citenamefont {Deichmann}, \citenamefont {van~der Vegt},\ and\ \citenamefont
  {Leroy}}]{Ardham2015}%
  \BibitemOpen
  \bibfield  {author} {\bibinfo {author} {\bibfnamefont {V.~R.}\ \bibnamefont
  {Ardham}}, \bibinfo {author} {\bibfnamefont {G.}~\bibnamefont {Deichmann}},
  \bibinfo {author} {\bibfnamefont {N.~F.}\ \bibnamefont {van~der Vegt}}, \
  and\ \bibinfo {author} {\bibfnamefont {F.}~\bibnamefont {Leroy}},\ }\bibfield
   {title} {\enquote {\bibinfo {title} {Solid-liquid work of adhesion of
  coarse-grained models of n-hexane on graphene layers derived from the
  conditional reversible work method},}\ }\href@noop {} {\bibfield  {journal}
  {\bibinfo  {journal} {J. Chem. Phys.}\ }\textbf {\bibinfo {volume} {143}},\
  \bibinfo {pages} {243135} (\bibinfo {year} {2015})}\BibitemShut {NoStop}%
\bibitem [{\citenamefont {Kandu{\v{c}}}\ and\ \citenamefont
  {Netz}(2017)}]{Kanduc2017}%
  \BibitemOpen
  \bibfield  {author} {\bibinfo {author} {\bibfnamefont {M.}~\bibnamefont
  {Kandu{\v{c}}}}\ and\ \bibinfo {author} {\bibfnamefont {R.~R.}\ \bibnamefont
  {Netz}},\ }\bibfield  {title} {\enquote {\bibinfo {title} {{Atomistic
  simulations of wetting properties and water films on hydrophilic
  surfaces}},}\ }\href {\doibase 10.1063/1.4979847} {\bibfield  {journal}
  {\bibinfo  {journal} {J Chem. Phys.}\ }\textbf {\bibinfo {volume} {146}},\
  \bibinfo {pages} {164705} (\bibinfo {year} {2017})}\BibitemShut {NoStop}%
\bibitem [{\citenamefont {Kandu{\v{c}}}(2017)}]{Kanduc2017a}%
  \BibitemOpen
  \bibfield  {author} {\bibinfo {author} {\bibfnamefont {M.}~\bibnamefont
  {Kandu{\v{c}}}},\ }\bibfield  {title} {\enquote {\bibinfo {title} {{Going
  beyond the standard line tension: Size-dependent contact angles of water
  nanodroplets}},}\ }\href {\doibase 10.1063/1.4990741} {\bibfield  {journal}
  {\bibinfo  {journal} {J. Chem. Phys.}\ }\textbf {\bibinfo {volume} {147}},\
  \bibinfo {pages} {174701} (\bibinfo {year} {2017})}\BibitemShut {NoStop}%
\bibitem [{\citenamefont {Jiang}, \citenamefont {M{\"u}ller-Plathe},\ and\
  \citenamefont {Panagiotopoulos}(2017)}]{Jiang2017}%
  \BibitemOpen
  \bibfield  {author} {\bibinfo {author} {\bibfnamefont {H.}~\bibnamefont
  {Jiang}}, \bibinfo {author} {\bibfnamefont {F.}~\bibnamefont
  {M{\"u}ller-Plathe}}, \ and\ \bibinfo {author} {\bibfnamefont {A.~Z.}\
  \bibnamefont {Panagiotopoulos}},\ }\bibfield  {title} {\enquote {\bibinfo
  {title} {Going beyond the standard line tension: Size-dependent contact
  angles of water nanodroplets},}\ }\href@noop {} {\bibfield  {journal}
  {\bibinfo  {journal} {J. Chem. Phys.}\ }\textbf {\bibinfo {volume} {147}},\
  \bibinfo {pages} {084708} (\bibinfo {year} {2017})}\BibitemShut {NoStop}%
\bibitem [{\citenamefont {Surblys}\ \emph {et~al.}(2018)\citenamefont
  {Surblys}, \citenamefont {Leroy}, \citenamefont {Yamaguchi},\ and\
  \citenamefont {M{\"u}ller-Plathe}}]{Surblys2018}%
  \BibitemOpen
  \bibfield  {author} {\bibinfo {author} {\bibfnamefont {D.}~\bibnamefont
  {Surblys}}, \bibinfo {author} {\bibfnamefont {F.}~\bibnamefont {Leroy}},
  \bibinfo {author} {\bibfnamefont {Y.}~\bibnamefont {Yamaguchi}}, \ and\
  \bibinfo {author} {\bibfnamefont {F.}~\bibnamefont {M{\"u}ller-Plathe}},\
  }\bibfield  {title} {\enquote {\bibinfo {title} {Molecular dynamics analysis
  of the influence of coulomb and van der waals interactions on the work of
  adhesion at the solid-liquid interface},}\ }\href {\doibase
  10.1063/1.3601055} {\bibfield  {journal} {\bibinfo  {journal} {J. Chem.
  Phys.}\ }\textbf {\bibinfo {volume} {148}},\ \bibinfo {eid} {134707}
  (\bibinfo {year} {2018})}\BibitemShut {NoStop}%
\bibitem [{\citenamefont {Ravipati}\ \emph {et~al.}(2018)\citenamefont
  {Ravipati}, \citenamefont {Aymard}, \citenamefont {Kalliadasis},\ and\
  \citenamefont {Galindo}}]{Ravipati2018}%
  \BibitemOpen
  \bibfield  {author} {\bibinfo {author} {\bibfnamefont {S.}~\bibnamefont
  {Ravipati}}, \bibinfo {author} {\bibfnamefont {B.}~\bibnamefont {Aymard}},
  \bibinfo {author} {\bibfnamefont {S.}~\bibnamefont {Kalliadasis}}, \ and\
  \bibinfo {author} {\bibfnamefont {A.}~\bibnamefont {Galindo}},\ }\bibfield
  {title} {\enquote {\bibinfo {title} {On the equilibrium contact angle of
  sessile liquid drops from molecular dynamics simulations},}\ }\href@noop {}
  {\bibfield  {journal} {\bibinfo  {journal} {J. Chem. Phys.}\ }\textbf
  {\bibinfo {volume} {148}},\ \bibinfo {pages} {164704} (\bibinfo {year}
  {2018})}\BibitemShut {NoStop}%
\bibitem [{\citenamefont {Giacomello}, \citenamefont {Schimmele},\ and\
  \citenamefont {Dietrich}(2016)}]{Giacomello2016}%
  \BibitemOpen
  \bibfield  {author} {\bibinfo {author} {\bibfnamefont {A.}~\bibnamefont
  {Giacomello}}, \bibinfo {author} {\bibfnamefont {L.}~\bibnamefont
  {Schimmele}}, \ and\ \bibinfo {author} {\bibfnamefont {S.}~\bibnamefont
  {Dietrich}},\ }\bibfield  {title} {\enquote {\bibinfo {title} {{Wetting
  hysteresis induced by nanodefects}},}\ }\href {\doibase
  10.1073/pnas.1513942113} {\bibfield  {journal} {\bibinfo  {journal} {Proc.
  Natl. Acad. Sci. U. S. A.}\ }\textbf {\bibinfo {volume} {113}},\ \bibinfo
  {pages} {E262--E271} (\bibinfo {year} {2016})}\BibitemShut {NoStop}%
\bibitem [{\citenamefont {Zhang}, \citenamefont {M{\"{u}}ller-Plathe},\ and\
  \citenamefont {Leroy}(2015)}]{Zhang2015}%
  \BibitemOpen
  \bibfield  {author} {\bibinfo {author} {\bibfnamefont {J.}~\bibnamefont
  {Zhang}}, \bibinfo {author} {\bibfnamefont {F.}~\bibnamefont
  {M{\"{u}}ller-Plathe}}, \ and\ \bibinfo {author} {\bibfnamefont
  {F.}~\bibnamefont {Leroy}},\ }\bibfield  {title} {\enquote {\bibinfo {title}
  {Pinning of the contact line during evaporation on heterogeneous surfaces:
  Slowdown or temporary immobilization? insights from a nanoscale study},}\
  }\href {\doibase 10.1021/acs.langmuir.5b01097} {\bibfield  {journal}
  {\bibinfo  {journal} {Langmuir}\ }\textbf {\bibinfo {volume} {31}},\ \bibinfo
  {pages} {7544--7552} (\bibinfo {year} {2015})}\BibitemShut {NoStop}%
\bibitem [{\citenamefont {Zhang}, \citenamefont {Huang},\ and\ \citenamefont
  {Lu}(2019)}]{Zhang2019}%
  \BibitemOpen
  \bibfield  {author} {\bibinfo {author} {\bibfnamefont {J.}~\bibnamefont
  {Zhang}}, \bibinfo {author} {\bibfnamefont {H.}~\bibnamefont {Huang}}, \ and\
  \bibinfo {author} {\bibfnamefont {X.~Y.}\ \bibnamefont {Lu}},\ }\bibfield
  {title} {\enquote {\bibinfo {title} {Pinning-depinning mechanism of the
  contact line during evaporation of nanodroplets on heated heterogeneous
  surfaces: A molecular dynamics simulation},}\ }\href {\doibase
  10.1021/acs.langmuir.9b00796} {\bibfield  {journal} {\bibinfo  {journal}
  {Langmuir}\ }\textbf {\bibinfo {volume} {35}},\ \bibinfo {pages} {6356--6366}
  (\bibinfo {year} {2019})}\BibitemShut {NoStop}%
\bibitem [{\citenamefont {Wilhelmy}(1863)}]{Wilhelmy1863}%
  \BibitemOpen
  \bibfield  {author} {\bibinfo {author} {\bibfnamefont {L.}~\bibnamefont
  {Wilhelmy}},\ }\bibfield  {title} {\enquote {\bibinfo {title} {{Ueber die
  Abh\"{a}ngigkeit der Capillarit\"{a}ts-Constanten des Alkohols von Substanz
  und Gestalt des benetzten festen K\"{o}rpers}},}\ }\href {\doibase
  10.1002/andp.18631950602} {\bibfield  {journal} {\bibinfo  {journal} {Ann.
  Phys.}\ }\textbf {\bibinfo {volume} {195}},\ \bibinfo {pages} {177--217}
  (\bibinfo {year} {1863})},\ \Eprint
  {http://arxiv.org/abs/https://onlinelibrary.wiley.com/doi/pdf/10.1002/andp.18631950602}
  {https://onlinelibrary.wiley.com/doi/pdf/10.1002/andp.18631950602}
  \BibitemShut {NoStop}%
\bibitem [{\citenamefont {Volpe}\ and\ \citenamefont
  {Siboni}(2018)}]{Volpe2018}%
  \BibitemOpen
  \bibfield  {author} {\bibinfo {author} {\bibfnamefont {C.~D.}\ \bibnamefont
  {Volpe}}\ and\ \bibinfo {author} {\bibfnamefont {S.}~\bibnamefont {Siboni}},\
  }\bibfield  {title} {\enquote {\bibinfo {title} {{The Wilhelmy method: a
  critical and practical review}},}\ }\href {\doibase 10.1680/jsuin.17.00059}
  {\bibfield  {journal} {\bibinfo  {journal} {Surf. Innov.}\ }\textbf {\bibinfo
  {volume} {6}},\ \bibinfo {pages} {120--132} (\bibinfo {year}
  {2018})}\BibitemShut {NoStop}%
\bibitem [{\citenamefont {Marchand}\ \emph {et~al.}(2012)\citenamefont
  {Marchand}, \citenamefont {Weijs}, \citenamefont {Snoeijer},\ and\
  \citenamefont {Andreotti}}]{Marchand2012}%
  \BibitemOpen
  \bibfield  {author} {\bibinfo {author} {\bibfnamefont {A.}~\bibnamefont
  {Marchand}}, \bibinfo {author} {\bibfnamefont {J.~H.}\ \bibnamefont {Weijs}},
  \bibinfo {author} {\bibfnamefont {J.~H.}\ \bibnamefont {Snoeijer}}, \ and\
  \bibinfo {author} {\bibfnamefont {B.}~\bibnamefont {Andreotti}},\ }\bibfield
  {title} {\enquote {\bibinfo {title} {Why is surface tension a force parallel
  to the interface?}}\ }\href@noop {} {\bibfield  {journal} {\bibinfo
  {journal} {Am. J. Phys.}\ }\textbf {\bibinfo {volume} {79}},\ \bibinfo
  {pages} {999--1008} (\bibinfo {year} {2012})}\BibitemShut {NoStop}%
\bibitem [{\citenamefont {Das}\ \emph {et~al.}(2011)\citenamefont {Das},
  \citenamefont {Marchand}, \citenamefont {Andreotti},\ and\ \citenamefont
  {Snoeijer}}]{Das2011}%
  \BibitemOpen
  \bibfield  {author} {\bibinfo {author} {\bibfnamefont {S.}~\bibnamefont
  {Das}}, \bibinfo {author} {\bibfnamefont {A.}~\bibnamefont {Marchand}},
  \bibinfo {author} {\bibfnamefont {B.}~\bibnamefont {Andreotti}}, \ and\
  \bibinfo {author} {\bibfnamefont {J.~H.}\ \bibnamefont {Snoeijer}},\
  }\bibfield  {title} {\enquote {\bibinfo {title} {Elastic deformation due to
  tangential capillary forces},}\ }\href@noop {} {\bibfield  {journal}
  {\bibinfo  {journal} {Phys. Fluids}\ }\textbf {\bibinfo {volume} {23}},\
  \bibinfo {pages} {1--11} (\bibinfo {year} {2011})}\BibitemShut {NoStop}%
\bibitem [{\citenamefont {Weijs}, \citenamefont {Andreotti},\ and\
  \citenamefont {Snoeijer}(2013)}]{Weijs2013}%
  \BibitemOpen
  \bibfield  {author} {\bibinfo {author} {\bibfnamefont {J.~H.}\ \bibnamefont
  {Weijs}}, \bibinfo {author} {\bibfnamefont {B.}~\bibnamefont {Andreotti}}, \
  and\ \bibinfo {author} {\bibfnamefont {J.~H.}\ \bibnamefont {Snoeijer}},\
  }\bibfield  {title} {\enquote {\bibinfo {title} {{Elasto-capillarity at the
  nanoscale: on the coupling between elasticity and surface energy in soft
  solids}},}\ }\href {\doibase 10.1039/c3sm50861g} {\bibfield  {journal}
  {\bibinfo  {journal} {Soft Matter}\ }\textbf {\bibinfo {volume} {9}},\
  \bibinfo {pages} {8494} (\bibinfo {year} {2013})}\BibitemShut {NoStop}%
\bibitem [{\citenamefont {Merchant}\ and\ \citenamefont
  {Keller}(1992)}]{Merchant1992}%
  \BibitemOpen
  \bibfield  {author} {\bibinfo {author} {\bibfnamefont {G.~J.}\ \bibnamefont
  {Merchant}}\ and\ \bibinfo {author} {\bibfnamefont {J.~B.}\ \bibnamefont
  {Keller}},\ }\bibfield  {title} {\enquote {\bibinfo {title} {Line tension
  between fluid phases and a substrate},}\ }\href {\doibase 10.1063/1.858320}
  {\bibfield  {journal} {\bibinfo  {journal} {Phys. Fluids A}\ }\textbf
  {\bibinfo {volume} {4}},\ \bibinfo {pages} {477} (\bibinfo {year}
  {1992})}\BibitemShut {NoStop}%
\bibitem [{\citenamefont {Getta}\ and\ \citenamefont
  {Dietrich}(1998)}]{Getta1998}%
  \BibitemOpen
  \bibfield  {author} {\bibinfo {author} {\bibfnamefont {T.}~\bibnamefont
  {Getta}}\ and\ \bibinfo {author} {\bibfnamefont {S.}~\bibnamefont
  {Dietrich}},\ }\bibfield  {title} {\enquote {\bibinfo {title} {Line tension
  between fluid phases and a substrate},}\ }\href {\doibase
  10.1103/PhysRevE.57.655} {\bibfield  {journal} {\bibinfo  {journal} {Phys.
  Rev. E}\ }\textbf {\bibinfo {volume} {57}},\ \bibinfo {pages} {655--671}
  (\bibinfo {year} {1998})}\BibitemShut {NoStop}%
\bibitem [{\citenamefont {Mastny}\ and\ \citenamefont
  {de~Pablo}(2007)}]{Mastny2007}%
  \BibitemOpen
  \bibfield  {author} {\bibinfo {author} {\bibfnamefont {E.~A.}\ \bibnamefont
  {Mastny}}\ and\ \bibinfo {author} {\bibfnamefont {J.~J.}\ \bibnamefont
  {de~Pablo}},\ }\bibfield  {title} {\enquote {\bibinfo {title} {{Melting line
  of the Lennard-Jones system, infinite size, and full potential}},}\ }\href
  {\doibase 10.1063/1.2753149} {\bibfield  {journal} {\bibinfo  {journal} {J.
  Chem. Phys.}\ }\textbf {\bibinfo {volume} {127}},\ \bibinfo {pages} {104504}
  (\bibinfo {year} {2007})}\BibitemShut {NoStop}%
\bibitem [{\citenamefont {Boruvka}\ and\ \citenamefont
  {Neumann}(1977)}]{Boruvka1977}%
  \BibitemOpen
  \bibfield  {author} {\bibinfo {author} {\bibfnamefont {L.}~\bibnamefont
  {Boruvka}}\ and\ \bibinfo {author} {\bibfnamefont {A.~W.}\ \bibnamefont
  {Neumann}},\ }\bibfield  {title} {\enquote {\bibinfo {title} {Generalization
  of the classical theory of capillarity},}\ }\href@noop {} {\bibfield
  {journal} {\bibinfo  {journal} {J. Chem. Phys.}\ }\textbf {\bibinfo {volume}
  {66}},\ \bibinfo {pages} {5464--5476} (\bibinfo {year} {1977})}\BibitemShut
  {NoStop}%
\bibitem [{\citenamefont {Marmur}(1997)}]{Marmur1997line}%
  \BibitemOpen
  \bibfield  {author} {\bibinfo {author} {\bibfnamefont {A.}~\bibnamefont
  {Marmur}},\ }\bibfield  {title} {\enquote {\bibinfo {title} {Line tension and
  the intrinsic contact angle in solid--liquid--fluid systems},}\ }\href@noop
  {} {\bibfield  {journal} {\bibinfo  {journal} {J. Colloid Interface Sci.}\
  }\textbf {\bibinfo {volume} {186}},\ \bibinfo {pages} {462--466} (\bibinfo
  {year} {1997})}\BibitemShut {NoStop}%
\bibitem [{\citenamefont {Thompson}\ \emph {et~al.}(1984)\citenamefont
  {Thompson}, \citenamefont {Gubbins}, \citenamefont {Walton}, \citenamefont
  {Chantry},\ and\ \citenamefont {Rowlinson}}]{Thompson1984}%
  \BibitemOpen
  \bibfield  {author} {\bibinfo {author} {\bibfnamefont {S.~M.}\ \bibnamefont
  {Thompson}}, \bibinfo {author} {\bibfnamefont {K.~E.}\ \bibnamefont
  {Gubbins}}, \bibinfo {author} {\bibfnamefont {J.~P. R.~B.}\ \bibnamefont
  {Walton}}, \bibinfo {author} {\bibfnamefont {R.~A.~R.}\ \bibnamefont
  {Chantry}}, \ and\ \bibinfo {author} {\bibfnamefont {J.~S.}\ \bibnamefont
  {Rowlinson}},\ }\bibfield  {title} {\enquote {\bibinfo {title} {A molecular
  dynamics study of liquid drops},}\ }\href@noop {} {\bibfield  {journal}
  {\bibinfo  {journal} {J. Chem. Phys.}\ }\textbf {\bibinfo {volume} {81}},\
  \bibinfo {pages} {530--542} (\bibinfo {year} {1984})}\BibitemShut {NoStop}%
\bibitem [{\citenamefont {Yaguchi}, \citenamefont {Yano},\ and\ \citenamefont
  {Fujikawa}(2010)}]{Yaguchi2010}%
  \BibitemOpen
  \bibfield  {author} {\bibinfo {author} {\bibfnamefont {H.}~\bibnamefont
  {Yaguchi}}, \bibinfo {author} {\bibfnamefont {T.}~\bibnamefont {Yano}}, \
  and\ \bibinfo {author} {\bibfnamefont {S.}~\bibnamefont {Fujikawa}},\
  }\bibfield  {title} {\enquote {\bibinfo {title} {Molecular dynamics study of
  vapor-liquid equilibrium state of an argon nanodroplet and its vapor},}\
  }\href {\doibase 10.1299/jfst.5.180} {\bibfield  {journal} {\bibinfo
  {journal} {J. Fluid. Sci. Tech.}\ }\textbf {\bibinfo {volume} {5}},\ \bibinfo
  {pages} {180} (\bibinfo {year} {2010})}\BibitemShut {NoStop}%
\bibitem [{\citenamefont {Irving}\ and\ \citenamefont
  {Kirkwood}(1950)}]{Irving1950}%
  \BibitemOpen
  \bibfield  {author} {\bibinfo {author} {\bibfnamefont {J.~H.}\ \bibnamefont
  {Irving}}\ and\ \bibinfo {author} {\bibfnamefont {J.~G.}\ \bibnamefont
  {Kirkwood}},\ }\bibfield  {title} {\enquote {\bibinfo {title} {{The
  statistical mechanical theory of transport processes. IV. The equations of
  hydrodynamics}},}\ }\href {\doibase 10.1063/1.1747782} {\bibfield  {journal}
  {\bibinfo  {journal} {J. Chem. Phys.}\ }\textbf {\bibinfo {volume} {18}},\
  \bibinfo {pages} {817} (\bibinfo {year} {1950})}\BibitemShut {NoStop}%
\bibitem [{\citenamefont {Schofield}\ and\ \citenamefont
  {Henderson}(1982)}]{Schofield1982}%
  \BibitemOpen
  \bibfield  {author} {\bibinfo {author} {\bibfnamefont {D.}~\bibnamefont
  {Schofield}}\ and\ \bibinfo {author} {\bibfnamefont {J.~R.}\ \bibnamefont
  {Henderson}},\ }\bibfield  {title} {\enquote {\bibinfo {title} {Statistical
  mechanics of inhomogeneous fluids},}\ }\href@noop {} {\bibfield  {journal}
  {\bibinfo  {journal} {Proc. R. Soc. Lond. A}\ }\textbf {\bibinfo {volume}
  {379}},\ \bibinfo {pages} {231--246} (\bibinfo {year} {1982})}\BibitemShut
  {NoStop}%
\bibitem [{\citenamefont {Rowlinson}(1993)}]{Rowlinson1993}%
  \BibitemOpen
  \bibfield  {author} {\bibinfo {author} {\bibfnamefont {J.~S.}\ \bibnamefont
  {Rowlinson}},\ }\bibfield  {title} {\enquote {\bibinfo {title} {Themodynamics
  of inhomogeneous systems},}\ }\href@noop {} {\bibfield  {journal} {\bibinfo
  {journal} {Pure Appl. Chem.}\ }\textbf {\bibinfo {volume} {65}},\ \bibinfo
  {pages} {873--882} (\bibinfo {year} {1993})}\BibitemShut {NoStop}%
\bibitem [{\citenamefont {Frenkel}\ and\ \citenamefont
  {Smit}(1996)}]{Frenkel2007}%
  \BibitemOpen
  \bibfield  {author} {\bibinfo {author} {\bibfnamefont {D.}~\bibnamefont
  {Frenkel}}\ and\ \bibinfo {author} {\bibfnamefont {B.}~\bibnamefont {Smit}},\
  }\href@noop {} {\emph {\bibinfo {title} {Understanding Molecular Simulation:
  From Algorithms to Applications}}}\ (\bibinfo  {publisher} {Academic Press},\
  \bibinfo {year} {1996})\ pp.\ \bibinfo {pages} {152--156}\BibitemShut
  {NoStop}%
\end{thebibliography}
%
%
\end{document}